\documentclass[a4paper,fleqn,usenatbib,useAMS]{mn2e}

\usepackage[applemac]{inputenc}
\usepackage[draft=false, colorlinks=true, allcolors=blue]{hyperref}
\usepackage{graphicx}	
\usepackage{amsmath}	
\usepackage{amssymb}	
\usepackage{siunitx}    

\newcommand{\subscr}[1]{\ensuremath{_{\mathrm{#1}}}}
\newcommand{\supscr}[1]{\ensuremath{^{\mathrm{#1}}}}

\DeclareSIUnit\h{\ensuremath{\mathit{h}}}
\DeclareSIUnit\hs{\ensuremath{\mathit{h}_{70}}}
\DeclareSIUnit\parsec{pc}
\DeclareSIUnit\Msun{\ensuremath{M_{\odot}}}
\DeclareSIUnit\Zsun{\ensuremath{Z_{\odot}}}
\DeclareSIUnit\yr{\ensuremath{yr}}
\DeclareSIUnit\erg{\ensuremath{erg}}

\newcommand\ion[2]{\text{#1\,\textsc{\lowercase{#2}}}}	

\newcommand{\mpc}{\si{\mega\parsec}}

\newcommand{\msun}{\si{\Msun}}
\newcommand{\mh}{\si{\h^{-1}.\msun}}

\newcommand{\impch}{\si{\h.\mega\parsec^{-1}}}
\newcommand{\emean}[1]{\left\langle #1 \right\rangle}
\newcommand{\mean}[1]{\bar{#1}}
\newcommand{\vect}[1]{\ensuremath{\mathbf{#1}}}
\newcommand{\diff}{\ensuremath{\mathrm{d}}}

\newcommand{\bah}{\textsc{bahamas}}
\newcommand{\owls}{OWLS}
\newcommand{\cowls}{cosmo-OWLS}
\newcommand{\lcdm}{\ensuremath{\Lambda}CDM}
\newcommand{\Om}{\ensuremath{\Omega_\mathrm{m}}}
\newcommand{\Ob}{\ensuremath{\Omega_\mathrm{b}}}

\newcommand{\Ol}{\ensuremath{\Omega_\Lambda}}

\newcommand{\rhom}{\ensuremath{\mean{\rho}_{\mathrm{m}}}}

\newcommand{\cb}{\ensuremath{\mathtt{cb}}}
\newcommand{\nocb}{\ensuremath{\mathtt{nocb}}}

\defcitealias{Zu2015}{ZM15}

\usepackage[T1]{fontenc}
\usepackage{ae,aecompl}

\usepackage{newtxtext,newtxmath}

\title[Impact of baryons on the power spectrum]{The impact of the
  observed baryon distribution in haloes on the total matter power
  spectrum}

\author[S.N.B. Debackere et al.]{
  Stijn N.B. Debackere\thanks{Contact e-mail:
    \href{mailto:debackere@strw.leidenuniv.nl}{debackere@strw.leidenuniv.nl}
  },
  Joop Schaye,
  Henk Hoekstra
  \\
  Leiden Observatory, Leiden University, PO Box 9513, NL-2300 RA
  Leiden, The Netherlands
  \\}

\date{Last updated --; in original form --}

\pubyear{2019}

\begin{document}\label{firstpage}
\pagerange{\pageref{firstpage}--\pageref{lastpage}}
\maketitle

\begin{abstract}
  {The interpretation of upcoming weak gravitational lensing surveys
    depends critically on our understanding of the matter power
    spectrum on scales $k < \SI{10}{\impch}$, where baryonic processes
    are important. We study the impact of galaxy formation processes
    on the matter power spectrum using a halo model that treats the
    stars and gas separately from the dark matter distribution. We use
    empirical constraints from X-ray observations (hot gas) and halo
    occupation distribution modelling (stars) for the baryons. Since
    X-ray observations cannot generally measure the hot gas content
    outside $r\subscr{500c}$, we vary the gas density profiles beyond
    this radius. Compared with dark matter only models, we find a
    total power suppression of $\SI{1}{\percent}$ ($\SI{5}{\percent}$)
    on scales $\SIrange{0.2}{1}{\impch}$ ($\SIrange{0.5}{2}{\impch}$),
    where lower baryon fractions result in stronger suppression. We
    show that groups of galaxies
    ($\num{e13} < m\subscr{500c} / (\si{\mh}) < \num{e14}$) dominate
    the total power at all scales $k \lesssim \SI{10}{\impch}$. We
    find that a halo mass bias of $\SI{30}{\percent}$ (similar to what
    is expected from the hydrostatic equilibrium assumption) results
    in an underestimation of the power suppression of up to
    $\SI{4}{\percent}$ at $k=\SI{1}{\impch}$, illustrating the
    importance of measuring accurate halo masses. Contrary to work
    based on hydrodynamical simulations, our conclusion that baryonic
    effects can no longer be neglected is not subject to uncertainties
    associated with our poor understanding of feedback processes.
    Observationally, probing the outskirts of groups and clusters will
    provide the tightest constraints on the power suppression for
    $k \lesssim \SI{1}{\impch}$.}
\end{abstract}




\section{Introduction}\label{sec:introduction}
Since the discovery of the Cosmic Microwave Background (CMB)
\citep{Penzias1965, Dicke1965}, cosmologists have continuously refined
the values of the cosmological parameters. This resulted in the
discovery of the accelerated expansion of the Universe
\citep{Riess1998, Perlmutter1999} and the concordance Lambda cold dark
matter (\lcdm) model. Future surveys such as
Euclid\footnote{\url{http://www.euclid-ec.org}}, the Large Synoptic
Survey Telescope (LSST)\footnote{\url{http://www.lsst.org/}}, and the
Wide Field Infra-Red Survey Telescope
(WFIRST)\footnote{\url{http://wfirst.gsfc.nasa.gov}} aim to constrain
the nature of this mysterious acceleration to establish whether it is
caused by a cosmological constant or dark energy. This is one of the
largest gaps in our current understanding of the Universe.

To probe the physical cause of the accelerated expansion, and to
discern between different models for dark energy or even a modified
theory of gravity, we require precise measurements of the growth of
structure and the expansion history over a range of redshifts. This is
exactly what future galaxy surveys aim to do, e.g. using a combination
of weak gravitational lensing and galaxy clustering. Weak lensing
measures the correlation in the distortion of galaxy shapes for
different redshift bins, which depends on the matter distribution in
the Universe, and thus on the matter power spectrum \citep[for
reviews, see e.g.][]{Hoekstra2008, Kilbinger2015, Mandelbaum2017}. The
theoretical matter power spectrum is thus an essential ingredient for
a correct interpretation of weak lensing observations.

The matter power spectrum can still not be predicted well at small
scales (\(k \gtrsim \SI{0.3}{\impch}\)) because of the uncertainty
introduced by astrophysical processes related to galaxy formation
\citep{Rudd2008, VanDaalen2011, Semboloni2011}. In order to provide
stringent cosmological constraints with future surveys, the prediction
of the matter power spectrum needs to be accurate at the sub-percent
level \citep{Hearin2012}.

Collisionless N-body simulations, i.e. dark matter only (DMO)
simulations, can provide accurate estimates of the non-linear effects
of gravitational collapse on the matter power spectrum. They can be
performed using a large number of particles, and in big cosmological
boxes for many different cosmologies (e.g. \citealp{Heitmann2009,
  Heitmann2010}; \citealp{Lawrence2010}; \citealp{Angulo2012}). The
distribution of baryons, however, does not perfectly trace that of the
dark matter: baryons can cool and collapse to high densities, sparking
the formation of galaxies. Galaxy formation results in violent
feedback that can redistribute gas to large scales. Furthermore, these
processes induce a back-reaction on the distribution of dark matter
\citep[e.g.][]{VanDaalen2011, VanDaalen2019, Velliscig2014}. Hence,
the redistribution of baryons and dark matter modifies the power
spectrum relative to that from DMO simulations.

Weak lensing measurements obtain their highest signal-to-noise ratio
on scales \(k \approx \SI{1}{\impch}\) (see \S 1.8.5 in
\citealp{Amendola2018}). \citet{VanDaalen2011} used the \owls\ suite
of cosmological simulations \citep{Schaye2010} to show that the
inclusion of baryon physics, particularly feedback from Active
Galactic Nuclei (AGN), influences the matter power spectrum at the
\SIrange{1}{10}{\percent} level between \(0.3 < k/(\si{\impch}) < 1\)
in their most realistic simulation that reproduced the hot gas
properties of clusters of galaxies. Further studies
\citep[e.g.][]{Vogelsberger2014a, Hellwing2016, Springel2017,
  Chisari2018, Peters2018, VanDaalen2019} have found similar results.
\citet{Semboloni2011} have shown, also using the \owls\ simulations,
that ignoring baryon physics in the matter power spectrum results in
biased weak lensing results, reaching a bias of up to
\SI{40}{\percent} in the dark energy equation of state parameter
\(w_{0}\) for a Euclid-like survey.

Current state-of-the-art hydrodynamical simulations allow us to study
the influence of baryons on the matter power spectrum, but cannot
predict it from first principles. Due to their computational cost,
these simulations need to include baryon processes such as star
formation and AGN feedback as ``subgrid'' recipes, as they cannot be
directly resolved. The accuracy of the subgrid recipes can be tested
by calibrating simulations to a fixed set of observed cosmological
scaling relations, and subsequently checking whether other scaling
relations are also reproduced \citep[see e.g.][]{Schaye2015,
  McCarthy2017, Pillepich2017a}. However, this calibration strategy
may not result in a unique solution, since other subgrid
implementations or different parameter values can provide similar
predictions for the calibrated relation but may differ in some other
observable. Thus, the calibrated relations need to be chosen carefully
depending on what we want to study.

A better option is to calibrate hydrodynamical simulations using the
observations that are most relevant for the power spectrum, such as
cluster gas fractions and the galaxy mass function
\citep{McCarthy2017} and to include simulations that span the
observational uncertainties \citep{McCarthy2018}. The calibration
against cluster gas fractions is currently only implemented in the
\bah\ suite of simulations \citep{McCarthy2017}. Current
high-resolution hydrodynamical simulations, such as e.g. EAGLE
\citep{Schaye2015}, Horizon-AGN \citep{Chisari2018} and IllustrisTNG
\citep{Springel2017}, do not calibrate against this observable.
Moreover, the calibrated subgrid parameters required to reproduce
their chosen observations result in gas fractions that are too high in
their most massive haloes \citep{Schaye2015, Barnes2017, Chisari2018}.
This is a problem, because both halo models \citep{Semboloni2013} and
hydrodynamical simulations \citep{VanDaalen2019} have been used to
demonstrate the existence of a strong link between the suppression of
the total matter power spectrum on large scales and cluster gas
fractions. As a result, these state-of-the-art simulations of galaxy
formation are not ideal to study the baryonic effects on the matter
power spectrum.

Focussing purely on simulation predictions risks underestimating the
possible range of power suppression due to baryons, since the
simulations generally do not cover the full range of possible physical
models. Hence, given our limited understanding of the astrophysics of
galaxy formation and the computational expense of hydrodynamical
simulations, it is important to develop other ways to account for
baryonic effects and observational constraints upon them.

One possibility is to make use of phenomenological models that take
the matter distribution as input without making assumptions about the
underlying physics. Splitting the matter into its dark matter and
baryonic components allows observations to be used as the input for
the baryonic component of the model. This bypasses the need for any
model calibrations but may require extrapolating the baryonic
component outside of the observed range. Such models can be
implemented in different ways. For instance, \citet{Schneider2015} and
\citet{Schneider2018} use a ``baryon correction model'' to shift the
particles in DMO simulations under the influence of hydrodynamic
processes which are subsumed in a combined density profile including
dark matter, gas and stars with phenomenological parameters for the
baryon distribution that are fit to observations. Consequently, the
influence of a change in these parameters on the power spectrum can be
investigated. Since this model relies only on DMO simulations, it is
less computationally expensive while still providing important
information on the matter distribution.

We will take a different phenomenological approach and use a modified
version of the halo model to predict how baryons modify the matter
power spectrum. We opt for this approach because it gives us freedom
in varying the baryon distribution at little computational expense. We
do not aim to make the most accurate predictor for baryonic effects on
the power spectrum, but our goal is to systematically study the
influence of changing the baryonic density profiles on the matter
power spectrum and to quantify the uncertainty of the baryonic effects
on the power spectrum allowed by current observational constraints.

The halo model describes the clustering of matter in the Universe
starting from the matter distribution of individual haloes. We split
the halo density profiles into a dark matter component and baryonic
components for the gas and the stars. We assume that the abundance and
clustering of haloes can be modelled using DMO simulations, but that
their density profiles, and hence masses, change due to baryonic
effects. This assumption is supported by the findings of
\citet{VanDaalen2014a}, who used \owls\ to show that matched sets of
subhaloes cluster identically on scales larger than the virial radii
in DMO and hydrodynamical simulations. We constrain the gas component
with X-ray observations of groups and clusters of galaxies. These
observations are particularly relevant since the matter power spectrum
is dominated by groups and clusters on the scales affected by baryonic
physics and probed by upcoming surveys
\(0.3 \lesssim k/(\impch) \lesssim 10\),
\citep[e.g.][]{VanDaalen2015}. For the stellar component, we assume
the distribution from Halo Occupation Distribution (HOD) modelling.

Earlier studies have used extensions to the halo model to include
baryon effects, either by adding individual matter components from
simulations \citep[e.g.][]{White2004, Zhan2004, Rudd2008, Guillet2010,
  Semboloni2011, Semboloni2013, Fedeli2014, Fedeli2014a}, or by
introducing empirical parameters inspired by the predicted physical
effects of galaxy formation \citep[see][]{Mead2015, Mead2016}.
However, these studies were based entirely on data from cosmological
simulations, whereas we stay as close as possible to the observations
and thus do not depend on the uncertain assumptions associated with
subgrid models for feedback processes.

There is still freedom in our model because the gas content of
low-mass haloes and the outskirts of clusters cannot currently be
measured. We thus study the range of baryonic corrections to the dark
matter only power spectrum by assuming different density profiles for
the unobserved regions. Our model gives us a handle on the uncertainty
in the matter power spectrum and allows us to quantify how different
mass profiles of different mass haloes contribute to the total power
for different wavenumbers, whilst simultaneously matching observations
of the matter distribution. Moreover, we can study the impact of
observational uncertainties and biases on the resulting power
spectrum.

We start of by describing our modified halo model in
\S~\ref{sec:hm}. We describe the observations and the relevant halo
model parameters in \S~\ref{sec:obs_xray}. We show our resulting
model density components in \S~\ref{sec:components} and report our
results in \S~\ref{sec:results}. We discuss our model and compare it
to the literature in \S~\ref{sec:discussion}. Finally, we conclude
and provide some directions for future research in
\S~\ref{sec:summary_conclusions}. This work assumes the \emph{WMAP}
9 year \citep{Hinshaw2013} cosmological parameters
\(\{\Om, \Ob, \Ol, \sigma_{8}, n\subscr{s}, h\} = \{0.2793, 0.0463,
0.7207, 0.821, 0.972, 0.7\}\) and all of our results are computed for
\(z=0\). All of the observations that we compare to assumed \(h=0.7\),
so we quote their results in units of
\(H_{0} = 70 \si{\hs.\km.\s^{-1}.\mpc^{-1}}\) with \(h_{70} = 1\).
Whenever we quote units without any \(\si{\h}\) or \(\si{\hs}\)
scaling, we assume \(\si{\h}=0.7\) or, equivalently, \(\si{\hs}=1\)
\citep[for a good reference and arguments on making definitions
explicit, see][]{Croton2013}. When fitting our model to observations,
we always use \(\si{\h}=0.7\) to ensure a fair comparison between
model and observations.

\section{Halo Model}\label{sec:hm}

\subsection{Theory}\label{sec:hm_theory}
The halo model (e.g. \citealp{Peacock2000,Seljak2000a}; but the basis
was already worked out in \citealp{McClelland1977a} and
\citealp{Scherrer1991}; review in \citealp{Cooray2002}) is an analytic
prescription to model the clustering properties of matter for a given
cosmology through the power spectrum \citep[for a clear pedagogical
exposition, see][]{Vandenbosch2013}. It gives insight into non-linear
structure formation starting from the linear power spectrum and a few
simplifying assumptions.

The spherical collapse model of non-linear structure formation tells
us that any over-dense, spherical region will collapse into a
virialized dark matter halo, with a final average density
\(\emean{\rho_{\mathrm{f}}} =
\Delta_{\mathrm{vir}}\rho_{\mathrm{c}}(z_{\mathrm{vir}})\), where
\(\Delta_{\mathrm{vir}}\) in general depends on cosmology, but is
usually taken as \(\Delta_{200}=200\), rounded from the Einstein-de
Sitter value of \(\Delta_{\mathrm{vir}} = 18 \pi^{2}\), with
\(\rho_{\mathrm{c}}(z_{\mathrm{vir}})\) the critical density of the
Universe at the redshift of virialization. The fundamental assumption
of the halo model is that all matter in the Universe has collapsed
into virialized dark matter haloes that grow hierarchically in time
through mergers. Throughout the paper we will adhere to the notation
\(m\subscr{500c}\) and \(m\subscr{200m}\) to indicate regions
enclosing an average density
\(\emean{\rho}\subscr{500c} = 500 \rho\subscr{c}(z)\) and
\(\emean{\rho}\subscr{200m} = 200 \rhom(z)\), with
$\rhom(z) = \Om \rho\subscr{c}(z=0) (1+z)^3$, respectively.

At a given time, the halo mass function \(n(m\subscr{h}, z)\)
determines the co-moving number density of dark matter haloes in a
given halo mass bin centered on $m\subscr{h}$. This function can be
derived from analytic arguments, like for instance the Press-Schechter
and Extended Press-Schechter (EPS) theories
\citep[e.g.][]{Press1974,Bond1991,Lacey1993}, or by using DMO
simulations \citep[e.g.][]{Sheth1999b, Jenkins2001, Tinker2008}.
Furthermore, assuming that the density profile of a halo is completely
determined by its mass and redshift, i.e.
\(\rho(r) = \rho(r|m\subscr{h}, z)\), we can then calculate the
statistics of the matter distribution in the Universe, captured by the
power spectrum, by looking at the correlations between matter in
different haloes (the two-halo or 2h term which probes large scales)
and between matter within the same halo (the one-halo or 1h term which
probes small scales).

Splitting the contributions to the power spectrum up into the 1h and
2h terms, we can rewrite
\begin{align}
  \label{eq:power}
  P(k,z) & = V\subscr{u} \emean{|\hat{\delta}\subscr{m}(k,z)|^{2}} \\
  \label{eq:power_12h}
       & = P\subscr{1h}(k,z) + P\subscr{2h}(k,z) \, .
\end{align}
Here \(V\subscr{u}\) is the volume under consideration and
\(\hat{\delta}\subscr{m}(k,z)\) is the Fourier transform of the matter
overdensity field
\(\delta\subscr{m}(\vect{x},z) \equiv \rho(\vect{x},z) / \rhom(z) -
1\), with $\rhom(z)$ the mean matter background density at redshift
$z$. We define the Fourier transform of a halo as
\begin{align}
  \label{eq:rho_k}
  \hat{\rho}(k|m\subscr{h}, z) & = 4\pi \int_0^{r\subscr{h}} \diff r
                                 \, \rho(r|m\subscr{h},
                                 z) r^2 \frac{\sin(kr)}{kr} \, .
\end{align}
The 1h and 2h terms are given by \citep[for detailed derivations, see
][]{Cooray2002,Mo2010}
\begin{align}
  \label{eq:p_1h}
  P\subscr{1h}(k,z) & = \int \diff m\subscr{200m,dmo} \,
                      \begin{aligned}[t]
                        & n\subscr{dmo}(m\subscr{200m,dmo}(z),z) \\
                        \times & \frac{|\hat{\rho}(k|m\subscr{h}(m\subscr{200m,dmo}),z)|^{2}}{\rhom^{2}(z)} \\
                    \end{aligned} \\
  \nonumber
  P\subscr{2h}(k,z) & = P_{\mathrm{lin}}(k,z)
                      \begin{aligned}[t]
                        \Bigg[ \int \diff
                        m\subscr{200m,dmo} \, &
                        n\subscr{dmo}(m\subscr{200m,dmo}(z),z) \\
                        \times & b\subscr{dmo}(m\subscr{200m,dmo}(z),z) \\
                        \times & \frac{\hat{\rho}(k|m\subscr{h}(m\subscr{200m,dmo}),z)}{\rhom(z)}
                        \Bigg]^{2}
                    \end{aligned} \\
  \label{eq:p_2h}
                  & \simeq P_{\mathrm{lin}}(k,z) \, .
\end{align}
Our notation makes explicit that because our predictions rely on the
halo mass function and the bias obtained from DMO simulations, we need
to correct the true halo mass \(m\subscr{h}\) to the DMO equivalent
mass \(m\subscr{200m,dmo}\), as we will explain further in
\S~\ref{sec:hm_modifications}. The 2h term contains the bias
$b\subscr{dmo}(m,z)$ between haloes and the underlying density field.
For the 2h term, we simply use the linear power spectrum, which we get
from CAMB\footnote{\url{http://camb.info/}} for our cosmological
parameters. For the halo mass function, we assume the functional form
given by \citet{Tinker2008}, which is calibrated for the spherical
overdensity halo mass $m\subscr{200m,dmo}$.

We assume $P\subscr{2h} \approx P\subscr{lin}$ since not all of our
haloes will be baryonically closed. This would result in
Eq.~\ref{eq:p_2h} not returning to the linear power spectrum at large
scales for models that have missing baryons within the halo radius.
Assuming that the 2h term follows the linear power spectrum is
equivalent to assuming that all of the missing baryons will be
accounted for in the cosmic web, which we cannot accurately capture
with our simple halo model.

We will use our model to predict the quantity
\begin{align}
  \label{eq:power_ratio}
    R_{i}(k,z) \equiv \frac{P_{i}(k,z)}{P_{i,\mathrm{dmo}}(k,z)} \,
  ,
\end{align}
the ratio between the power spectrum of baryonic model $i$ and the
corresponding DMO power spectrum assuming the same cosmological
parameters. This ratio has been given various names in the literature,
e.g. the ``response'' \citep{Mead2016}, the ``reaction''
\citep{Cataneo2018}, or just the ``suppression''
\citep{Schneider2018}. We will refer to it as the power spectrum
response to the presence of baryons. It quantifies the suppression or
increase of the matter power spectrum due to baryons. If non-linear
gravitational collapse and galaxy formation effects were separable,
and baryonic effects were insensitive to the underlying cosmology,
knowledge of this ratio would allow us to reconstruct a matter power
spectrum from any DMO prediction. These last two assumptions can only
be tested by comparing large suites of cosmological N-body and
hydrodynamical simulations. We do not attempt to address them in this
paper. However, \citet{VanDaalen2011, VanDaalen2019},
\citet{Mummery2017}, \citet{McCarthy2018}, and \citet{Stafford2019}
have investigated the cosmology dependence of the baryonic
suppression. \citet{Mummery2017} find that a separation of the
cosmology and baryon effects on the power spectrum is accurate at the
$\SI{3}{\percent}$ level between
$\SI{1}{\impch} \lesssim k \lesssim \SI{10}{\impch}$ for cosmologies
varying the neutrino masses between $0 < M_\nu / \si{\eV} < 0.48$.
Similarly, \citet{VanDaalen2019} find that varying the cosmology
between \emph{WMAP} 9 and Planck 2013 results in at most a
$\SI{4}{\percent}$ difference for $k < \SI{10}{\impch}$.

Our model does not include any correction to the power spectrum due to
halo exclusion. Halo exclusion accounts for the fact that haloes
cannot overlap by canceling the 2h term at small scales
\citep{Smith2011a}. It also cancels the shot-noise contribution from
the 1h term at large scales. In our model, the important effect occurs
at scales where the 1h and 2h terms are of similar magnitude, since
the halo exclusion would suppress the 2h term. However, since we look
at the power spectrum response to baryons $R_i(k)$, which is the ratio
of the power spectrum including baryons to the power spectrum in the
DMO case, our model should not be significantly affected, since the
halo exclusion term modifies both of these terms in a similar way. We
have checked that subtracting a halo exclusion term that interpolates
between the 1h term at large scales and the 2h term at small scales
only affects our predictions for $R_i(k)$ by at most
$\SI{1}{\percent}$ at $k \approx \SI{3}{\impch}$.

\subsection{Linking observed halo masses to abundances}\label{sec:hm_modifications}
Our model is similar to the traditional halo model as described by
\citet{Cooray2002}. We make two important changes, however. Firstly,
we split up the density profile into a dark matter, a hot gas, and a
stellar component
\begin{align}
  \label{eq:rho_r}
  \rho(r|m\subscr{h}, z) & = \rho\subscr{dm}(r|m\subscr{h}, z) +
                           \rho\subscr{gas}(r|m\subscr{h}, z) +
                           \rho\subscr{\star}(r|m\subscr{h}, z) \, .
\end{align}
We will detail our specific profile assumptions in
\S~\ref{sec:hm_profiles}. Secondly, we include a mapping from the
observed halo mass $m\subscr{h}$ to the dark matter only equivalent
halo mass $m\subscr{200m,dmo}$, as shown in Eqs.~\ref{eq:p_1h} and
\ref{eq:p_2h}.

This second step is necessary for two reasons. First, the masses of
haloes change in hydrodynamical simulations. In simulations with the
same initial total density field, haloes can be linked between the
collisionless and hydrodynamical simulations, thus enabling the study
of the impact of baryon physics on individual haloes.
\citet{Sawala2013}, \citet{Velliscig2014} and \citet{Cui2014} found
that even though the abundance of individual haloes does not change,
their mass does, especially for low-mass haloes (see Fig. 10 in
\citealp{Velliscig2014}). Feedback processes eject gas from haloes,
lowering their mass at fixed radius. However, once this mass change is
accounted for, the clustering of the matched haloes is nearly
identical in the DMO and hydrodynamical simulations
\citep{VanDaalen2014a}. Since the halo model relies on prescriptions
for the halo mass function that are calibrated on dark matter only
simulations, we need to correct our observed halo masses to predict
their abundance.

Second, observed halo masses are not equivalent to the underlying true
halo mass. Every observational determination of the halo mass carries
its own intrinsic biases. Masses from X-ray measurements are generally
obtained under the assumption of spherical symmetry and hydrostatic
equilibrium, for example. However, due to the recent assembly of
clusters of galaxies, sphericity and equilibrium assumptions break
down in the halo outskirts \citep[see][and references
therein]{Pratt2019}. In most weak lensing measurements, the halo is
modeled assuming a Navarro-Frenk-White (NFW) profile
\citep{Navarro1996} with a concentration-mass relation $c(m)$ from
simulations. This profile does not necessarily accurately describe the
density profile of individual haloes due to asphericity and the large
scatter in the concentration-mass relation at fixed halo mass.

In our model, each halo will be labeled with four different halo
masses. We indicate the cumulative mass profile of the observed and
DMO equivalent halo with $m\subscr{obs}(\leq r)$ and
$m\subscr{dmo}(\leq r)$, respectively. Firstly, we define the total
mass inside $r\subscr{500c,obs}$ inferred from observations
\begin{align}
  \label{eq:m500c_xray}
  m\subscr{500c,obs} & \equiv
                       m\subscr{obs}(\leq r\subscr{500c,obs}) \, .
\end{align}
This mass will provide the link between our model and the
observations. We work with $r\subscr{500c,obs}$ in this paper because
it is similar to the radius up to which X-ray observations are able to
measure the halo mass. However, any other radius can readily be used
in all of the following definitions. Secondly, we have the true total
mass inside the halo radius $r\subscr{h}$ for our extrapolated
profiles
\begin{align}
  \label{eq:m_h}
  m\subscr{h} & \equiv m\subscr{obs}(\leq r\subscr{h}) \, .
\end{align}
Thirdly, we define the total mass in our extrapolated profiles such
that the mean enclosed density is \(\emean{\rho}\subscr{200m}\)
\begin{align}
  \label{eq:m200m_obs}
  m\subscr{200m,obs} & \equiv m\subscr{obs}(\leq r\subscr{200m,obs}) \, .
\end{align}
We differentiate between $r\subscr{h}$ and $r\subscr{200m,obs}$
because for some of our models we will extrapolate the density profile
further than $r\subscr{200m,obs}$. Fourthly, we define the dark matter
only equivalent mass for the halo
\begin{align}
  \label{eq:m200m_dmo}
  m\subscr{200m,dmo} & \equiv m\subscr{dmo}(\leq
                       r\subscr{200m,dmo}(m\subscr{500c,obs},
                       c\subscr{dmo}(m\subscr{200m,dmo}))) \, ,
\end{align}
which depends on the observed halo mass $m\subscr{500c,obs}$ and the
assumed DMO concentration-mass relation
$c\subscr{dmo}(m\subscr{200m,dmo})$, as we will discuss below. In each
of our models for the baryonic matter distribution there is a unique
monotonic mapping between all four of these halo masses. In the rest
of the paper we will thus express all dependencies as a function of
$m\subscr{h}$, unless our calculation explicitly depends on one of the
three other masses (as we indicate in Eqs.~\ref{eq:p_1h} and
\ref{eq:p_2h} where the halo mass function requires the DMO equivalent
mass from Eq.~\ref{eq:m200m_dmo} as an input).

The DMO equivalent mass, Eq.~\ref{eq:m200m_dmo}, requires more
explanation. We determine it from the following, simplifying but
overall correct, assumption: the inclusion of baryon physics does not
significantly affect the distribution of the dark matter. This
assumption is corroborated by the findings of \citet{Duffy2010},
\citet{Velliscig2014} and \citet{Schaller2015}, who all find that in
hydrodynamical simulations that are able to reproduce many observables
related to the baryon distribution, the baryons do not significantly
impact the dark matter distribution. This assumption breaks down on
galaxy scales where the dark matter becomes more concentrated due to
the condensation of baryons at the center of the halo. However, these
scales are smaller than the scales of interest for upcoming weak
lensing surveys. Moreover, at these scales the stellar component
typically dominates over the dark matter. Assuming that the dark
matter component will have the same scale radius as its DMO equivalent
halo, we can convert the observed halo mass into its DMO equivalent.
The first step is to compute the dark matter mass in the observed
halo,
\begin{align}
  \label{eq:m500c_dm}
  m\subscr{500c,dm} & = m\subscr{500c,obs}
                      \begin{aligned}[t]
                        (1 & - f\subscr{gas,500c,obs}(m\subscr{500c,obs}) \\
                        & -
                        f\subscr{\star,500c,obs}(m\subscr{200m,dmo}(m\subscr{500c,obs}))
                        \, .
                       \end{aligned}
\end{align}
The dark matter mass is obtained by subtracting the observed gas and
stellar mass inside \(r\subscr{500c,obs}\) from the observed total
halo mass. The stellar fraction depends on the DMO equivalent halo
mass since we take the stellar profiles from the \texttt{iHOD} model
by \citet[][hereafter \citetalias{Zu2015}]{Zu2015}, which also uses a
halo model that is based on the \citet{Tinker2008} halo mass function.
This requires us to iteratively solve for the DMO equivalent mass
\(m\subscr{200m,dmo}\). Next, we assume that the DMO equivalent halo
mass at the radius \(r\subscr{500c,obs}\) is given by
\(m\subscr{500c,dmo} = (1 - \Ob / \Om)^{-1} m\subscr{500c,dm}\), which
is consistent with our assumption that baryons do not change the
distribution of dark matter. Subsequently, we can determine the halo
mass \(m\subscr{200m,dmo}\) by assuming a DMO concentration-mass
relation, an NFW density profile, and solving
$m\subscr{dmo}(\leq
r\subscr{500c,obs};c\subscr{dmo}(m\subscr{200m,dmo})) =
m\subscr{500c,dmo}$ for $m\subscr{200m,dmo}$. Thus, we determine
$m\subscr{200m,dmo}$ (Eq.~\ref{eq:m200m_dmo}) by solving the following
equation:
\begin{align}
  \label{eq:m200m_dmo_constraints}
  & 4 \pi \int_0^{r\subscr{500c,obs}}
    \rho\subscr{NFW}(r;c\subscr{dmo}(m\subscr{200m,dmo}(m\subscr{500c,obs})))
    r^{2} \diff r \\
  \nonumber
  & = \frac{m\subscr{500c,obs}}{1 - \Ob / \Om}
     \begin{aligned}[t]
       (1 & - f\subscr{gas,500c,obs}(m\subscr{500c,obs}) \\
       & -
       f\subscr{\star,500c,obs}(m\subscr{200m,dmo}(m\subscr{500c,obs})))
       \, .
     \end{aligned}
\end{align}
We determine the stellar fraction at $r\subscr{500c,obs}$ by assuming
the stellar profiles detailed in \S~\ref{sec:hm_profiles_stars}.
Finally, we obtain the relation
\(m\subscr{200m,dmo}(m\subscr{500c,obs})\) that assigns a DMO
equivalent mass to each observed halo with mass
\(m\subscr{500c,obs}\).

We initiate our model on an equidistant log-grid of halo masses
\(\SI{e10}{\mh} \leq m\subscr{500c,obs} \leq \SI{e15}{\mh}\), which we
sample with 101 bins. We show that our results are converged with
respect to our chosen mass range and binning in
App.~\ref{app:results_mass_range}. For each halo mass, we get the DMO
equivalent mass \(m\subscr{200m,dmo}\), the stellar fraction
\(f_{\star,i}(m\subscr{200m,dmo})\), with
\(i \in \{\mathrm{cen,sat}\}\), and the concentration of the DMO
equivalent halo $c\subscr{dmo}(m\subscr{200m,dmo})$. We will specify
all of our different matter component profiles in
\S~\ref{sec:hm_profiles}.

\subsection{Matter density profiles}\label{sec:hm_profiles}
In this section, we give the functional forms of the density profiles
that we use in our halo model. We assume three different matter
components: dark matter, gas and stars. The dark matter and stellar
profiles are taken directly from the literature, whereas we obtain the
gas profiles by fitting to observations from the literature. In our
model, we only include the hot, X-ray emitting gas with
$T > \SI{e7}{\K}$, thus neglecting the interstellar medium (ISM)
component of the gas. The ISM component is confined to the scale of
individual galaxies, where it can provide a similar contribution to
the total baryonic mass as the stars. The only halo masses for which
the total baryonic mass of the galaxy may be similar to that of the
surrounding diffuse circum-galactic medium (CGM) are Milky Way-like
galaxies, or even lower-mass haloes
\citep{Catinella2010,Saintonge2011}. However, these do not contribute
significantly to the total power at our scales of interest, as we will
show in \S~\ref{sec:results_masses}.

\subsubsection{Hot gas}\label{sec:hm_profiles_gas}
For the density profiles of hot gas, we assume traditionally used beta
profiles \citep{Cavaliere1978} inside \(r\subscr{500c,obs}\) where we
have observational constraints. We will extrapolate the beta profile
as a power-law with slope $-\gamma$ outside $r\subscr{500c,obs}$. In
our models with $r\subscr{h} > r\subscr{200m,obs}$, we will assume a
constant density outside $r\subscr{200m,obs}$ until $r\subscr{h}$,
which will then be the radius where the halo reaches the cosmic baryon
fraction. This results in the following density profile for the hot
gas:
\begin{equation}
  \label{eq:beta_gas}
  \rho\subscr{gas}(r|m\subscr{h}) =
  \begin{cases}
    \rho_{0}\left(1 + (r/r\subscr{c})^{2}\right)^{-3\beta/2}, & r <
    r\subscr{500c,obs} \\
    \rho\subscr{500c,obs}
    \left(\frac{r}{r\subscr{500c,obs}}\right)^{-\gamma},
    & r\subscr{500c,obs} \leq r < r\subscr{200m,obs} \\
    \rho\subscr{500c,obs}
    \left(\frac{r\subscr{200m,obs}}{r\subscr{500c,obs}}\right)^{-\gamma},
    & r\subscr{200m,obs} \leq r < r\subscr{h} \\
    0, & r \geq r\subscr{h} \, .
  \end{cases}
\end{equation}
The normalisation \(\rho_{0}\) is determined by the gas fractions
inferred from X-ray observations and normalises the profile to
\(m\subscr{gas,500c,obs}\) at \(r\subscr{500c,obs}\):
\begin{align}
  \nonumber
  \rho_{0} & = \frac{m\subscr{gas,500c,obs}}{4/3 \pi r\subscr{500c,obs}^{3}
             {}_{2}F_{1}(3/2,3\beta/2;5/2;-(r\subscr{500c,obs}/r\subscr{c})^{2})}
  \\
  \label{eq:rho_0_gas}
           & = \frac{500\rho\subscr{c} f\subscr{gas,500c,obs}(m\subscr{500c,obs})}{
             {}_{2}F_{1}(3/2,3\beta/2;5/2;-(r\subscr{500c,obs}/r\subscr{c})^{2})}
             \, .
\end{align}
Here \({}_{2}F_{1}(a,b;c;d)\) is the Gauss hypergeometric function.
The values for the core radius \(r\subscr{c}\), the slope \(\beta\),
and the hot gas fraction $f\subscr{gas,500c,obs}(m\subscr{500c,obs})$
are obtained by fitting observations, as we explain in
\S~\ref{sec:obs_xray}. The outer power-law slope $\gamma$ is in
principle a free parameter of our model, but as we explain below, it
is constrained by the total baryon content of the halo. We choose a
parameter range of $0 \leq \gamma \leq 3$.

For each halo, we determine $r\subscr{200m,obs}$ by determining the
mean enclosed density for the total mass profile (i.e. dark matter,
hot gas and stars). In the most massive haloes, a large part of the
baryons is already accounted for by the observed hot gas profile. As a
result, we need to assume a steep slope in these systems, since
otherwise their baryon fraction would exceed the cosmic one before
$r\subscr{200m,obs}$ is reached. Since the parameters of both the dark
matter and the stellar components are fixed, the only way to prevent
this is by setting a maximum value for the slope $-\gamma$ once the
observational best-fit parameters for the hot gas profile have been
determined. For each $\rho(r|m\subscr{h})$ we can calculate the value
of $\gamma$ such that the cosmic baryon fraction is reached at
$r\subscr{200m,obs}$. This will be the limiting value and only equal
or steeper slopes will be allowed. We will show the resulting
$\gamma(m\subscr{500c,obs})$-relation in \S~\ref{sec:components},
since it depends on the best-fit density profile parameters from the
observations that we will describe in \S~\ref{sec:obs_xray}. Being
the only free parameter in our model, $\gamma$ provides a clear
connection to observations. Deeper observations that can probe further
into the outskirts of haloes, can thus be straightforwardly
implemented in our model.

We will look at two different cases for the size of the haloes,
motivated by the observed hot gas fractions in \S~\ref{sec:obs_xray}
and by the lack of observational constraints outside
$r\subscr{500c,obs}$. We aim to include enough freedom in the halo
outskirts such that the actual baryon distribution will be encompassed
by the models. In both cases, we leave the power-law slope $\gamma$
free outside $r\subscr{500c,obs}$. The models differ outside
$r\subscr{200m,obs}$ since there are no firm observational constraints
on the extent of the baryonic distribution around haloes. In the first
case, we will truncate the power-law as soon as $r\subscr{200m,obs}$
is reached, thus enforcing $r\subscr{h} = r\subscr{200m,obs}$. This
corresponds to the halo definition that is used by \citet{Tinker2008}
in constructing their halo mass function. For the least massive haloes
in our model, this will result in haloes that are missing a
significant fraction of their baryons at $r\subscr{200m,obs}$, with
lower baryon fractions $f\subscr{bar,200m,obs}$ for steeper slopes,
i.e. higher values of $\gamma$. Since we assume the linear power
spectrum for the 2h term, we will still get the clustering predictions
on the large scales right. We will denote this case with the
quantifier \nocb, since the cosmic baryon fraction
$f\subscr{b}=\Ob/\Om$ is not reached for most haloes in this case. In
the second case, we will set
$r\subscr{h} = r\subscr{f\subscr{b}} > r\subscr{200m,obs}$ such that
all haloes reach the cosmic baryon fraction at $r\subscr{h}$, we will
denote this case with the quantifier \cb.

The \nocb\ and the \cb\ cases for each $\gamma$ result in the same
halo mass $m\subscr{200m,obs}$, since they only differ for
$r > r\subscr{200m,obs}$. Thus, they have the same DMO equivalent halo
mass and the same abundance $n(m\subscr{200m,dmo}(m))$ in
Eq.~\ref{eq:p_1h}. The difference between the two models is the
normalization and the shape of the Fourier density profile
$\hat{\rho}(k|m)$ which depends on the total halo mass $m\subscr{h}$
and the distribution of the hot gas. The halo mass $m\subscr{h}$ will
be higher in the \cb\ case due to the added baryons between
$r\subscr{200m,obs} < r < r\subscr{h}$, resulting in more power from
the 1h term. Since the baryons in the \cb\ case are added outside
$r\subscr{200m,obs}$ there will also be an increase in power on larger
scales.

For our parameter range $0 \leq \gamma \leq 3$, the \nocb\ and \cb\
cases encompass the possible power suppression in the Universe. For
massive systems, we have observational constraints on the total baryon
content inside $r\subscr{500c,obs}$ and our model variations capture
the possible variation in the outer density profiles. The distribution
of the baryons in the hot phase outside $r\subscr{500c,obs}$ is not
known observationally. However, it most likely depends on the halo
mass. For the most massive haloes, Sunyaev-Zel'dovich (SZ)
measurements of the hot baryons indicate that most baryons are
accounted for inside
$5 \, r\subscr{500c,obs} \approx 2 \, r\subscr{200m,obs}$
\citep[e.g.][]{Planck2013, LeBrun2015}. This need not be the case for
lower-mass systems where baryons can be more easily ejected out to
even larger distances. Moreover, there are also baryons that never
make it into haloes and that are distributed on large, linear scales.
The main uncertainty in the power suppression at large scales stems
from the baryonic content of the low-mass systems. The 1h term of
low-mass haloes becomes constant for $k \lesssim \SI{1}{\impch}$.
Hence, on large scales we capture the extreme case where the low-mass
systems retain no baryons (\nocb\ and $\gamma=3$) and all the missing
halo baryons are distributed on large, linear scales in the cosmic
web. We can also capture the other extreme where the low-mass systems
retain all of their baryons in the halo outskirts (\cb\ and
$\gamma=0$), since the details of the density profile do not matter on
scales $k < \SI{1}{\impch}$. Thus, the matter distribution in the
Universe will lie somewhere in between these two extremes captured by
our model.

\subsubsection{Dark matter}\label{sec:hm_profiles_dm}
We assume that the dark matter follows a Navarro-Frenk-White (NFW)
profile \citep{Navarro1996} with the concentration determined by the
\(c\subscr{200c,dmo}(m\subscr{200c,dmo}(m\subscr{500c,obs}))\)
relation from \citet{Correa2015c}, which is calculated using
\texttt{commah}\footnote{\url{https://github.com/astroduff/commah}},
assuming Eq.~\ref{eq:m200m_dmo} to get the DMO equivalent mass. We
assume a unique $c(m)$ relation with no scatter. We discuss the
influence of shifting the concentration-mass relation within its
scatter in App.~\ref{app:results_concentration}.

The concentration in \texttt{commah} is calculated with respect to
\(r\subscr{200c,dmo}\) (the radius where the average enclosed density
of the halo is \(200 \, \rho\subscr{c}\)), so we convert the
concentration to our halo definition by multiplying by the factor
\(r\subscr{200m,dmo}/r\subscr{200c,dmo}\) (for the DMO equivalent
halo). This needs to be solved iteratively for haloes with different
concentration $c\subscr{200m,dmo}(m\subscr{200m,dmo})$, since for each
input mass $m\subscr{200c,dmo}$ and resulting concentration
$c\subscr{200c,dmo}$, we need to find the corresponding
$m\subscr{200m,dmo}$ to convert $c\subscr{200c,dmo}$ to
$c\subscr{200m,dmo}$. We thus have for the dark matter component in
Eq.~\ref{eq:rho_r}
\begin{equation}
  \label{eq:nfw_dm}
  \rho\subscr{dm}(r|m\subscr{h}) =
  \begin{cases}
    \frac{m\subscr{x}}{4 \pi r
      \subscr{x}^{3}}\frac{c\subscr{x}^{3}}{Y(c\subscr{x})} \left(
      \frac{c\subscr{x}r}{r\subscr{x}} \right)^{-1} \left(1 +
      \frac{c\subscr{x}r}{r\subscr{x}}\right)^{-2}, & r \leq r\subscr{h} \\
    0, & r > r\subscr{h} \, .
  \end{cases}
\end{equation}
The halo radius $r\subscr{h}$ depends on the hot gas density profile
and is either $r\subscr{h}=r\subscr{200m,obs}$ in the case \nocb, or
$r\subscr{h}=r\subscr{f\subscr{b}}$, the radius where the cosmic
baryon fraction is reached, in the case \cb. We define
\(Y(c\subscr{x}) = \log(1 + c\subscr{x}) - c\subscr{x}/(1 +
c\subscr{x})\) and the concentration
\(c\subscr{x} = r\subscr{x}/r\subscr{s}\) with the scale radius
$r\subscr{s}$ indicating the radius at which the NFW profile has
logarithmic slope $-2$. The subscript `x' indicates the radius at
which the concentration is calculated, e.g. $\mathrm{x=200m}$. All of
the subscripted variables are a function of the halo mass
$m\subscr{500c,obs}$. The normalization factor in our definition
ensures that the NFW profile has mass $m\subscr{x}$ at radius
$r\subscr{x}$. For the dark matter component in our baryonic model, we
require the mass at $r\subscr{500c,obs}$ to equal the dark matter
fraction of the total observed mass $m\subscr{500c,obs}$
\begin{align}
  \label{eq:mx_dm}
  m\subscr{x} &= m\subscr{500c,obs}
                \begin{aligned}[t]
                  (1 & - f\subscr{gas,500c,obs}(m\subscr{500c,obs}) \\
                  & -
                  f\subscr{\star,500c,obs}(m\subscr{200m,dmo}(m\subscr{500c,obs})))
                  \, .
                \end{aligned}
\end{align}
We require the scale radius for the dark matter to be the same as the
scale radius of the equivalent DMO halo, thus
\begin{align}
  \label{eq:c500c_dm}
  c\subscr{x} & = c\subscr{200m,dmo}(m\subscr{200m,dmo}(m\subscr{500c,obs})) \cdot
                \frac{r\subscr{500c,obs}}{r\subscr{200m,dmo}} \, .
\end{align}
For the DMO power spectrum that we compare to in
Eq.~\ref{eq:power_ratio}, we assume $\mathrm{x=200m,dmo}$ in
Eq.~\ref{eq:nfw_dm} and we use both the halo mass and the
concentration derived for Eq.~\ref{eq:m200m_dmo}. The halo radius for
the dark matter only case is the same as in the corresponding baryonic
model. This is the logical choice since this means that in the case
where our model accounts for all of the baryons inside $r\subscr{h}$,
the DMO halo and the halo including baryons will have the same total
mass, only the matter distributions will be different. In the case
where not all the baryons are accounted for, we can then see the
influence on the power spectrum of baryons missing from the haloes.

\subsubsection{Stars}\label{sec:hm_profiles_stars}
For the stellar contribution we do not try to fit density profiles to
observations. We opt for this approach since it allows for a clear
separation between centrals and satellites. Moreover, it provides the
possibility of a self-consistent framework that is also able to fit
the galaxy stellar mass function and the galaxy clustering. Our model
can be straightforwardly modified to take stellar fractions and
profiles from observations, as we did for the hot gas. We implement
stars similarly to HOD methods, specifically the \texttt{iHOD} model
by \citetalias{Zu2015}. We will assume their stellar-to-halo mass
relations for both centrals and satellites. The \texttt{iHOD} model
can reproduce the clustering and lensing of a large sample of SDSS
galaxies spanning 4 decades in stellar mass by self-consistently
modelling the incompleteness of the observations. Moreover, the model
independently predicts the observed stellar mass functions. In our
case, since we have assumed a different cosmology, these results will
not necessarily be reproduced. However, we have checked that shifting
the halo masses at fixed abundance between the cosmology of
\citetalias{Zu2015} and ours only results in relative shifts of the
stellar mass fractions of $\approx \SI{10}{\percent}$ at fixed halo
mass.

We split up the stellar component into centrals and satellites
\begin{equation}
  \label{eq:rho_star}
  \rho\subscr{\star}(r|m\subscr{h}) =
  \rho\subscr{cen}(r|m\subscr{h}) + \rho\subscr{sat}(r|m\subscr{h}) \, .
\end{equation}
The size of typical central galaxies in groups and clusters is much
smaller than our scales of interest, so we can safely assume them to
follow delta profile density distributions, as is done in
\citetalias{Zu2015}
\begin{equation}
  \label{eq:delta_cen}
  \rho\subscr{cen}(r|m\subscr{h}) = f\subscr{cen,200m,dmo}(m\subscr{200m,dmo}) \,
  m\subscr{200m,dmo} \, \delta\supscr{D}(\vect{r})
  \, ,
\end{equation}
here \(f\subscr{cen}(m)\) is taken directly from the \texttt{iHOD} fit
and \(\delta\supscr{D}(\vect{r})\) is the Dirac delta function.

For the satellite galaxies, we assume the same profile as
\citetalias{Zu2015} and put the stacked satellite distribution at
fixed halo mass in an NFW profile
\begin{equation}
  \label{eq:nfw_sat}
  \rho\subscr{sat}(r|m\subscr{h}) =
  \begin{cases}
    \frac{m\subscr{x}}{4 \pi r
      \subscr{x}^{3}}\frac{c\subscr{x}^{3}}{Y(c\subscr{x})} \left(
      \frac{c\subscr{x}r}{r\subscr{x}} \right)^{-1} \left(1 +
      \frac{c\subscr{x}r}{r\subscr{x}}\right)^{-2}, & r \leq r\subscr{h} \\
    0, & r > r\subscr{h} \, ,
  \end{cases}
\end{equation}
which is the same NFW definition as Eq.~\ref{eq:nfw_dm}. The profile
also becomes zero for $r>r\subscr{h}$. Clearly, there will still be
galaxies outside of this radius in the Universe. However, in the
halo-based picture, we need to truncate the halo somewhere. Since the
stellar contribution is always subdominant to the gas and the dark
matter at the largest scales, we can safely truncate the profiles
without affecting our predictions at the percent level. We will take
our reference values in Eq.~\ref{eq:nfw_sat} at $\mathrm{x=200m,dmo}$.
As in \citetalias{Zu2015}, the satellites are less concentrated than
the parent dark matter halo by a factor $0.86$
\begin{align}
  \label{eq:m200m_sat}
  m\subscr{x} & = f\subscr{sat}(m\subscr{200m,dmo}) \, m\subscr{200m,dmo} \\
  \nonumber
  c\subscr{x} & = f\subscr{c,sat} \,
                c\subscr{200m,dmo}(m\subscr{200m,dmo})\\
  \label{eq:c200m_sat}
              & = 0.86 \,
                c\subscr{200m,dmo}(m\subscr{200m,dmo})\, .
\end{align}
We take the stellar fraction from the best fit model of
\citetalias{Zu2015}.

This less concentrated distribution of satellites is also found in
observations for massive systems in the local Universe
\citep{Lin2004,Budzynski2012,VanderBurg2015a}. However, the
observations generally find a concentration of
$c\subscr{sat} \approx \numrange{2}{3}$ for group and cluster mass
haloes, which is about a factor 2 lower than the dark matter
concentration. Similar results are found in the \bah\ simulations
\citep{McCarthy2017}. In low-mass systems, on the other hand, the
satellites tend to track the underlying dark matter profile quite
closely \citep{Wang2014a} with \(c\subscr{sat}(m) \approx c(m)\). The
value of \(f\subscr{c,sat} = 0.86\) is thus a good compromise between
these two regimes. We have checked that assuming $f\subscr{c,sat} = 1$
results in differences $< \SI{0.03}{\percent}$ at all $k$, with the
maximum difference reached at $k \approx \SI{30}{\impch}$.

\section{X-ray observations}\label{sec:obs_xray}
\begin{figure}
  \centering
  \includegraphics[width=\columnwidth]{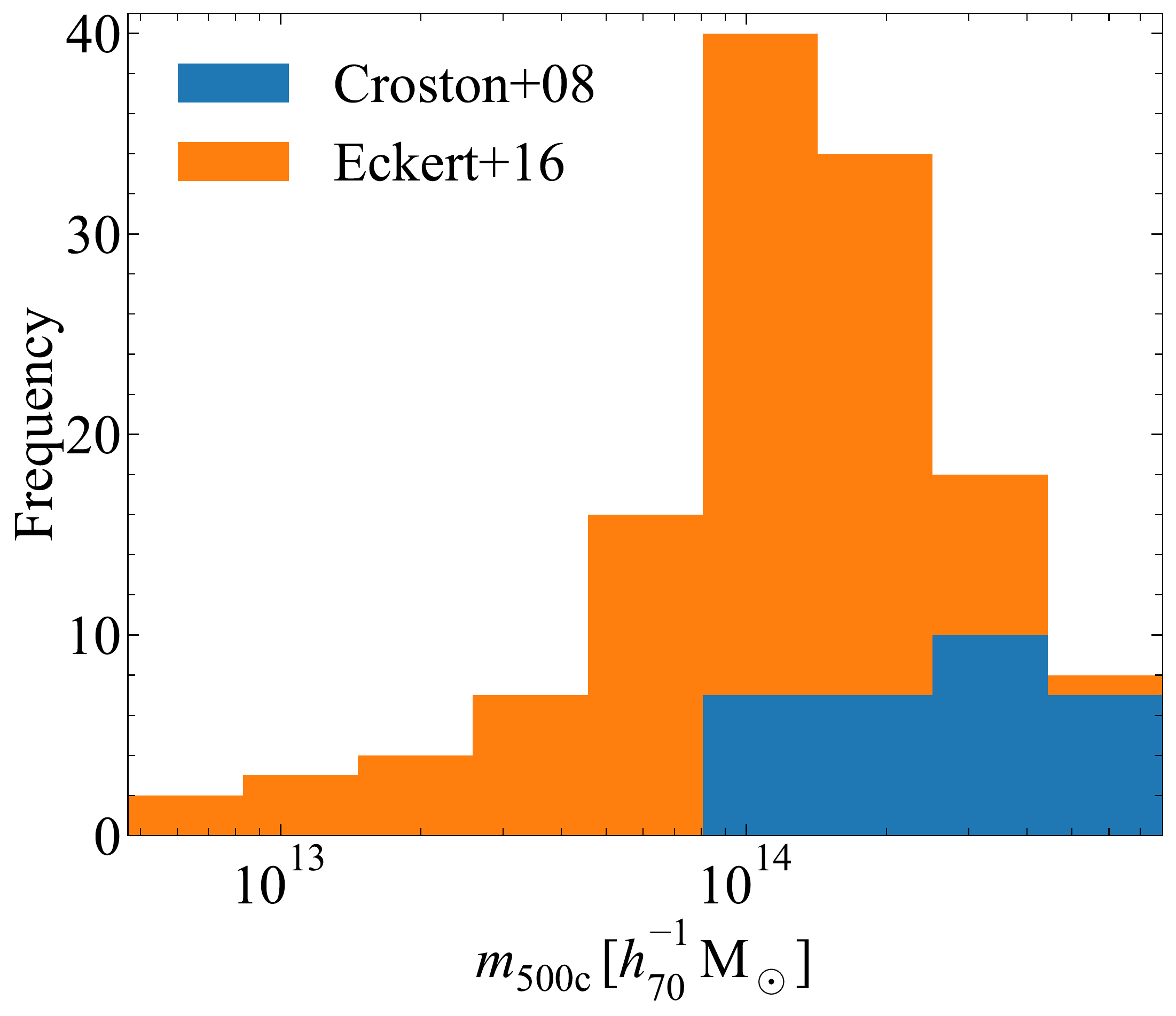}
  \caption{Stacked histogram for the masses of the haloes in our
    sample. The XXL-100-GC \citep{Eckert2016} data probe lower masses
    than the REXCESS \citep{Croston2008} data set, but it is clear
    that most of the haloes are clusters of galaxies with
    \(m\subscr{500c,obs} >
    \SI{e14}{\msun}\).}\label{fig:obs_masses_hist}
\end{figure}
\begin{figure}
  \centering
  \includegraphics[width=\columnwidth]{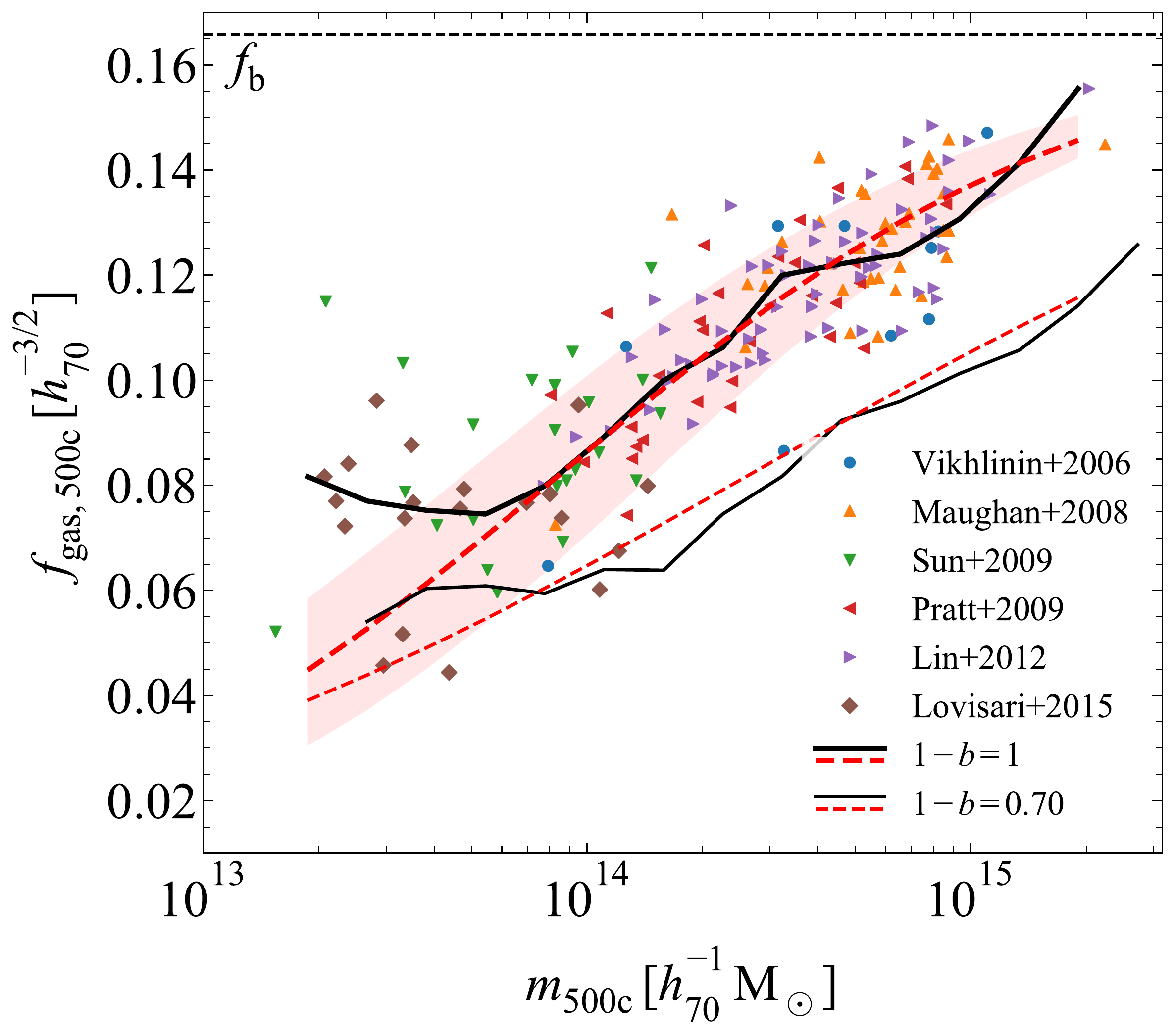}
  \caption{The X-ray hydrostatic gas fractions as a function of halo
    mass. The different data sets are explained in the text. The
    median \(f_{\mathrm{gas,500c}}-m_{\mathrm{500c,obs}}\) relation
    (black, solid lines) and the best fit (red, dashed lines) using
    Eq.~\ref{eq:fgas_sigmoid} are shown. We indicate the 15\supscr{th}
    and 85\supscr{th} percentile range by the red shaded region. We
    show the hydrostatic (thick lines) and bias corrected ($1-b=0.7$,
    thin lines) relations. Since, in the latter case, halo masses
    increase more than the gas masses, under the assumption of the
    best-fit beta profile to the hot gas density profiles, the gas
    fractions shift down. The fits deviate at low masses because we
    force \(f_{\mathrm{gas,500c}} \to 0\) for
    \(m_{\mathrm{500c,obs}} \to 0\).}\label{fig:obs_fgas}
\end{figure}
\begin{figure*}
  \centering
  \includegraphics[width=\textwidth]{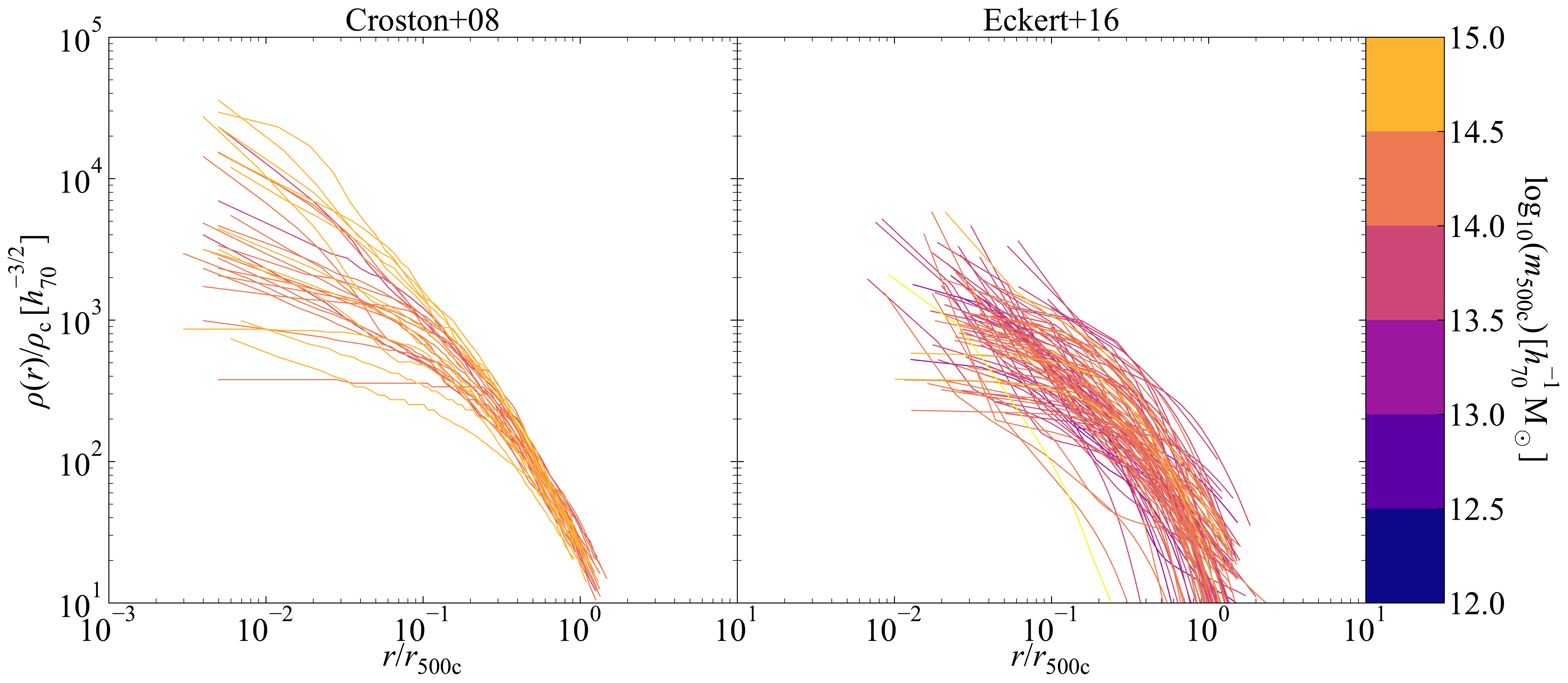}
  \caption{The hot gas density profiles inferred from the X-ray
    observations colour-coded by $m\subscr{500c,obs}$. The left-hand
    panel shows the profiles from the REXCESS sample \citep[31 nearby
    clusters with
    \(\num{e14} \lesssim m\subscr{500c,obs}/\si{\msun} \lesssim
    \num{e15}\), ][]{Croston2008} and the right-hand panel from the
    XXL survey \citep[100 bright clusters with
    \(\num{e13} \lesssim m\subscr{500c,obs}/\si{\msun} \lesssim
    \num{e15}\), ][]{Eckert2016}.}\label{fig:obs_profiles}
\end{figure*}
\begin{figure*}
  \centering
  \includegraphics[width=\textwidth]{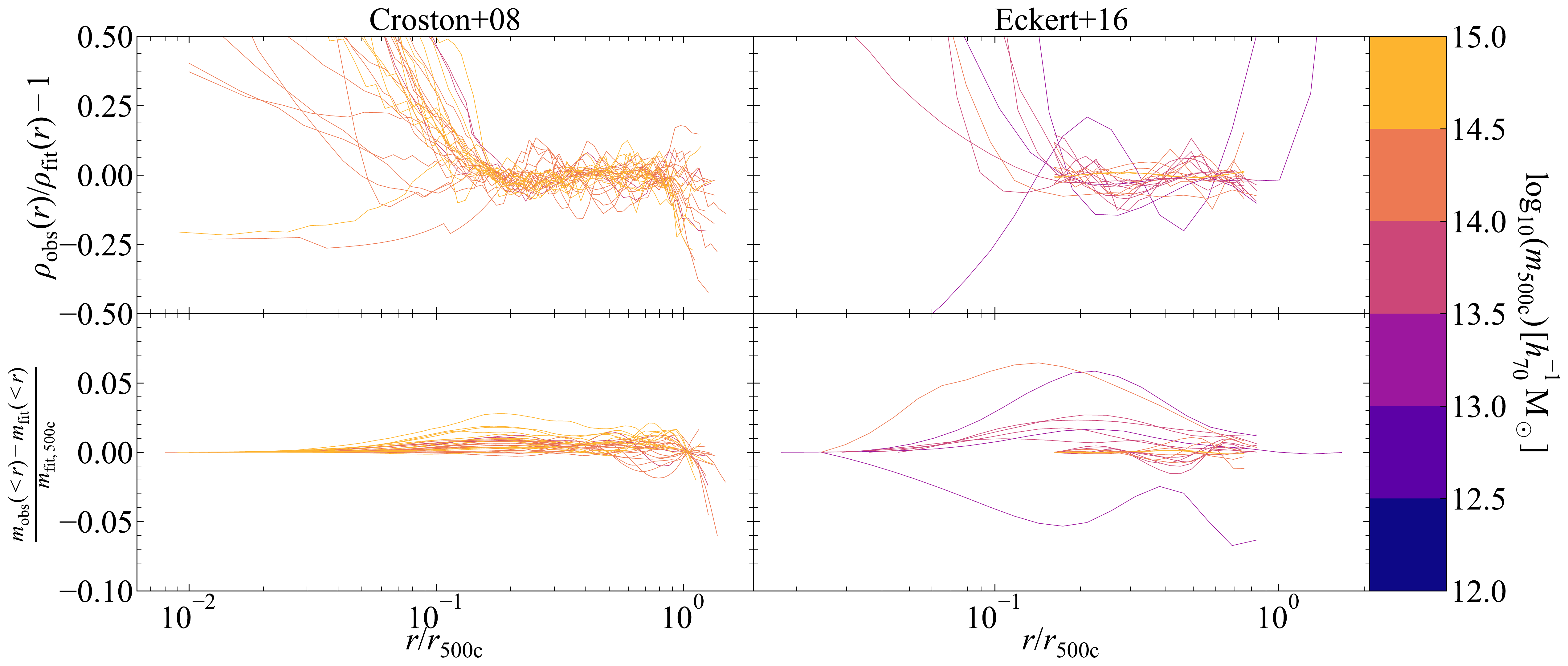}
  \caption{\emph{Top row} Residual of beta profile fits to the hot
    gas density profiles in Fig.~\ref{fig:obs_profiles}. The left-hand
    panel shows the residuals for the REXCESS sample
    \citep{Croston2008} and the right-hand panel for the XXL-100-GC
    sample \citep{Eckert2016}. The fits are accurate within
    \(\sim \SI{10}{\percent}\) for the range
    \(0.1 < r/r_{\mathrm{500c,obs}} < 1\), with larger scatter in the
    inner region, where the beta profile generally underestimates the
    total mass. For the XXL-100-GC sample \citep{Eckert2016} we binned
    the profiles into 20 mass bins since there is a large scatter in
    the individual ones. We only fit the profile at radial ranges
    where there is data for all the individual profiles in the bin.
    \emph{Bottom row} Residual of beta profile fits to the
    cumulative mass fraction. The total amount of mass within the
    inner region is negligible compared to the total mass in the
    profile. In the inner regions the observed profiles always yield
    higher masses than the fits because of the core of the beta
    profile. We reproduce the total mass \(m\subscr{gas,500c}\) of the
    individual density profiles by construction.}\label{fig:obs_fits}
\end{figure*}
We choose to constrain the halo model using observations of the hot,
X-ray emitting gas in groups and clusters of galaxies, since these
objects provide the dominant contribution to the power spectrum at our
scales of interest and their baryon content is dominated by hot
plasma.

We combine two data sets of X-ray observations with \emph{XMM-Newton}
of clusters for which the individually measured electron density
profiles were available, namely REXCESS \citep{Croston2008} and the
XXL survey \citep[more specifically the XXL-100-GC subset,
][]{Pacaud2016,Eckert2016}. This gives a total of 131 (\(31+100\))
unique groups and clusters (there is no overlap between the two data
sets) with masses ranging from
\(m\subscr{500c,obs} \approx \SI{1e13}{\msun}\) to
\(m\subscr{500c,obs} \approx \SI{2e15}{\msun}\), with the XXL sample
probing lower masses, as can be seen in
Fig.~\ref{fig:obs_masses_hist}. We extend our data with more sets of
observations for the hydrostatic gas fraction of groups and clusters
of galaxies, as shown in Fig.~\ref{fig:obs_fgas}. We use a set of
hydrostatic masses determined from \emph{Chandra} archival data
\citep{Vikhlinin2006b,Maughan2008b,Sun2009,Lin2012a} and from the
NORAS and REFLEX (of which REXCESS is a subset) surveys
\citep{Pratt2009,Lovisari2015}.

REXCESS consists of a representative sample of clusters from the
REFLEX survey \citep{Bohringer2007}. It includes clusters of all
dynamical states and aims to provide a homogeneous sampling in X-ray
luminosity of clusters in the local Universe (\(z < 0.2\)). Since all
of the redshift bins are approximately volume limited
\citep{Bohringer2007}, we do not expect significant selection effects
for the massive systems ($m\subscr{500c} > \SI{e14}{\mh}$) as it has
been shown by \citet{Chon2017} that the lack of disturbed clusters in
X-ray samples \citep{Eckert2011} is generally due to their
flux-limited nature. The XXL-100-GC sample is flux-limited
\citep{Pacaud2016} and covers a wider redshift range
(\(z \lesssim 1\)). Since it is flux-limited, there is a bias to
selecting more massive objects. At low redshifts, however, there is a
lack of massive objects due to volume effects \citep{Pacaud2016}. From
\citet{Chon2017} we would also expect the sample to be biased to
select relaxed systems.

Assuming an optically thin, collisionally-ionized plasma with a
temperature \(T\) and metallicity \(Z\), the deprojected surface
brightness profile can be converted into a 3-D electron density
profile \(n_{e}\), which is the source of the thermal bremsstrahlung
emission \citep{Sarazin1986}. For the REXCESS sample, the
spectroscopic temperature within \(r_{\mathrm{500c,obs}}\) was chosen
with the metallicity also deduced from a spectroscopic fit, whereas
for the XXL sample the average temperature within \(r<\SI{300}{kpc}\)
was used with a metallicity of \(Z= \SI{0.3}{\Zsun}\). We get the
corresponding hydrogen and helium abundances by interpolating between
the sets of primordial abundances,
\((X_{0},Y_{0},Z_{0}) = (0.75, 0.25, 0)\), and of solar abundances,
\((\mathrm{X_{\odot}},\mathrm{Y_{\odot}},\mathrm{Z_{\odot}})=(0.7133,0.2735,0.0132)\).
We then find \((X,Y,Z)=(0.73899,0.25705,0.00396)\) for
\(Z=\SI{0.3}{\Zsun}\). To convert this electron density into the total
density, we will assume these interpolated abundances, since in
general for clusters the metallicity
\(Z \approx \SI{0.3}{\Zsun} = 0.00396\) \citep{Voit2005b,
  Grevesse2007}. This is also approximately correct for the
\citet{Croston2008} data, since for their systems the median
metallicity (bracketed by 15\supscr{th} and 85\supscr{th} percentiles)
is \(Z/\si{\Zsun}=0.27\substack{+0.09 \\ -0.05}\). Moreover, we assume
the gas to be fully ionized. We know that the total gas density is
given by
\begin{align}
  \nonumber
  \rho_{\mathrm{gas}} & = \mu m_{\ion{H}{}} (n_{e} + n_{\ion{H}{}} + n_{\ion{He}{}})\\
  \nonumber
                      & = \frac{1+Y/X}{2+3Y/(4X)} \frac{2 + 3Y/(4X)}{1
                        + Y/(2X)} m_{\ion{H}{}} n_{e}\\
  \label{eq:ne_to_rho}
                      & \approx 0.6 \cdot 1.93 \, m_{\ion{H}{}} n_{e}
\end{align}
This results in the gas density profiles shown in
Fig.~\ref{fig:obs_profiles}. It is clear that at large radii the
scatter is smaller for more massive systems. We bin the XXL data in 20
mass bins as the individual profiles have a large scatter at fixed
radius. For each mass bin we only include the radial range where each
profile in the bin is represented.

The two surveys derived the halo mass $m\subscr{500c,obs}$
differently. For REXCESS, the halo masses for the whole sample were
determined from the \(m_{\mathrm{500c,obs}}-Y\subscr{X}\) relation of
\citet{Arnaud2007}, where
$Y\subscr{X} = m\subscr{gas,500c} \, T\subscr{X}$ is the thermal
energy content of the intracluster medium (ICM). \citet{Arnaud2007}
determined \(m_{\mathrm{500c,obs}}\) under the assumption of spherical
symmetry and hydrostatic equilibrium \citep[see e.g.][]{Voit2005b}.
\citet{Eckert2016} take a different route. They determine halo masses
using the \(m_{\mathrm{500c,obs}}-T\subscr{X}\) relation calibrated to
weak lensing mass measurements of 38 clusters that overlap with the
CFHTLenS shear catalog, as described in \citet{Lieu2016}. As a result,
the REXCESS halo mass estimates rely on the assumption of hydrostatic
equilibrium, whereas \citet{Eckert2016} actually find a hydrostatic
bias \(m\subscr{X-ray}/m\subscr{WL} = 1-b=0.72\), consistent with the
analyses of \cite{VonderLinden2014} and \cite{Hoekstra2015}. Recently,
\citet{Umetsu2019} used the Hyper Suprime-Cam (HSC) survey shear
catalog, which overlaps the XXL-North field almost completely, to
measure weak lensing halo masses with a higher limiting magnitude and,
hence, number density of source galaxies than CHFTLenS. They do not
rederive the gas fractions of \citet{Eckert2016}, but they note that
their masses are systematically lower by a factor $\approx 0.75$ than
those derived in \citet{Lieu2016}, a finding which is consistent with
\citet{Lieu2017}, who find a factor $\approx 0.72$. These lower weak
lensing halo masses result in a hydrostatic bias of $b < 0.1$.

To obtain a consistent analysis, we scale the halo masses from
\citet{Eckert2016} back onto the hydrostatic
\(f\subscr{gas,500c}-m\subscr{500c,obs}\) relation, which we show in
Fig.~\ref{fig:obs_fgas}. We thus assume halo masses derived from the
assumption of hydrostatic equilibrium. It might seem strange to take
the biased result as the starting point of our analysis. However, we
argue that this is an appropriate starting point. First, current
estimates for the hydrostatic bias range from
$0.58 \pm 0.04 \lesssim 1-b \lesssim 0.71 \pm 0.10$ corresponding to
the results from Planck SZ cluster counts
\citep{PlanckXXIV2016,Zubeldia2019}, or
$0.688\pm 0.072 \lesssim 1-b \lesssim 0.80 \pm 0.14$ from weak lensing
mass measurements of Planck clusters
\citep{VonderLinden2014,Hoekstra2015,Medezinski2018}. Second, we are
not able to determine the mass dependence of the relation for groups
of galaxies from current observations. We will check how our results
change when assuming a constant hydrostatic bias of \(1-b=0.7\) in
\S~\ref{sec:results_bias}. The thin, black line in
Fig.~\ref{fig:obs_fgas} shows the shift in the
$f\subscr{gas,500c}(m\subscr{500c,obs})$ relation when assuming this
constant hydrostatic bias.

We fit the cluster gas density profiles with beta profiles, following
Eq.~\ref{eq:beta_gas}, within \([0.15,1]\,r_{\mathrm{500c,obs}}\),
excising the core as usual in the literature, since it can deviate
from the flat slope in the beta profile. In observations, it is common
to assume a sum of different beta profiles to capture the slope in the
inner \(0.15 r\subscr{500c,obs}\). However, we correct for the mass
that we miss in the core by fixing the normalization to reproduce the
total gas mass of the profile, which is captured by the gas fraction
\(f\subscr{gas,500c}\). (This is equivalent to redistributing the
small amount of mass that we would miss in the core to larger scales.)
The slope at large radii, \(\beta\), and the core radius,
\(r\subscr{c}\), are the final two parameters determining the profile.
We show the residuals of the profile fits in Fig.~\ref{fig:obs_fits}
where we also include the residuals of the cumulative mass profile. It
is clear from the residuals in the top panels of
Fig.~\ref{fig:obs_fits} that the beta profile cannot accurately
capture the inner density profile of the hot gas. \citet{Arnaud2010}
show that the inner slope can vary from shallow to steep in going from
disturbed to relaxed or cool-core clusters. This need not concern us
because the deviations from the fit occur at such small radii that
they will not be able to significantly affect the power at our scales
of interest where the normalization of \(\hat{\rho}\subscr{gas}(k)\)
and, thus, the total mass of the hot gas component is the important
parameter. In the bottom panel of Fig.~\ref{fig:obs_fits} we show the
residuals for the cumulative mass. The left-hand panel of the figure
clearly shows that we force \(m\subscr{gas,500c}\) in the individual
profiles to equal the observed mass.

\begin{figure}
  \centering
  \includegraphics[width=\columnwidth]{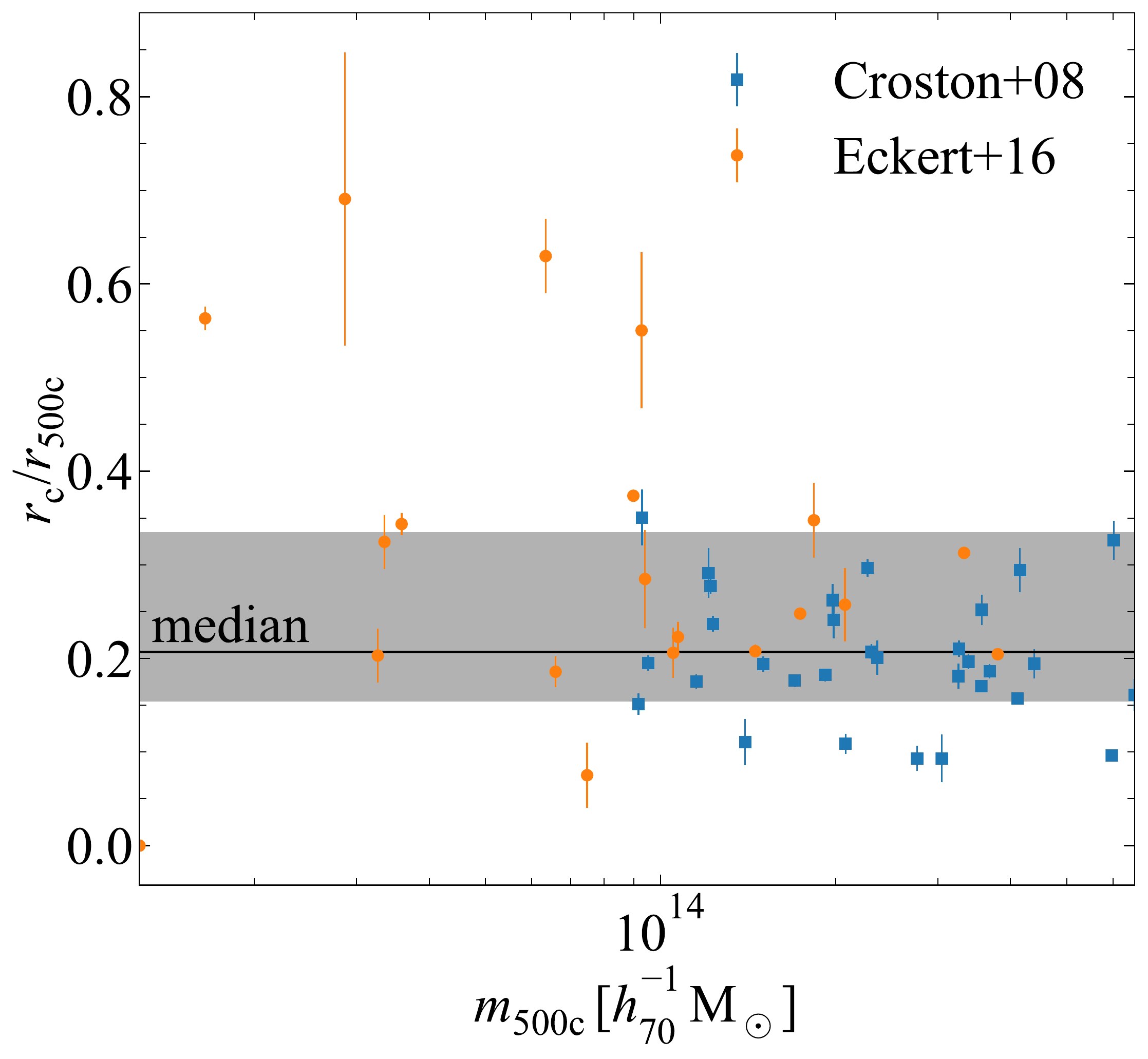}
  \caption{The mass dependence of the core radius $r\subscr{c}$ of the
    beta hot gas density profile fits, Eq.~\ref{eq:beta_gas}. We
    indicate the 15\supscr{th} and 85\supscr{th} percentiles with the
    gray shaded region and the median by the solid line. The error
    bars indicate the standard deviation in the best-fit parameter. We
    have binned the \citet{Eckert2016} sample into 20 mass bins. There
    is no clear mass dependence.}\label{fig:obs_rc_fit}
\end{figure}
\begin{figure}
  \centering
  \includegraphics[width=\columnwidth]{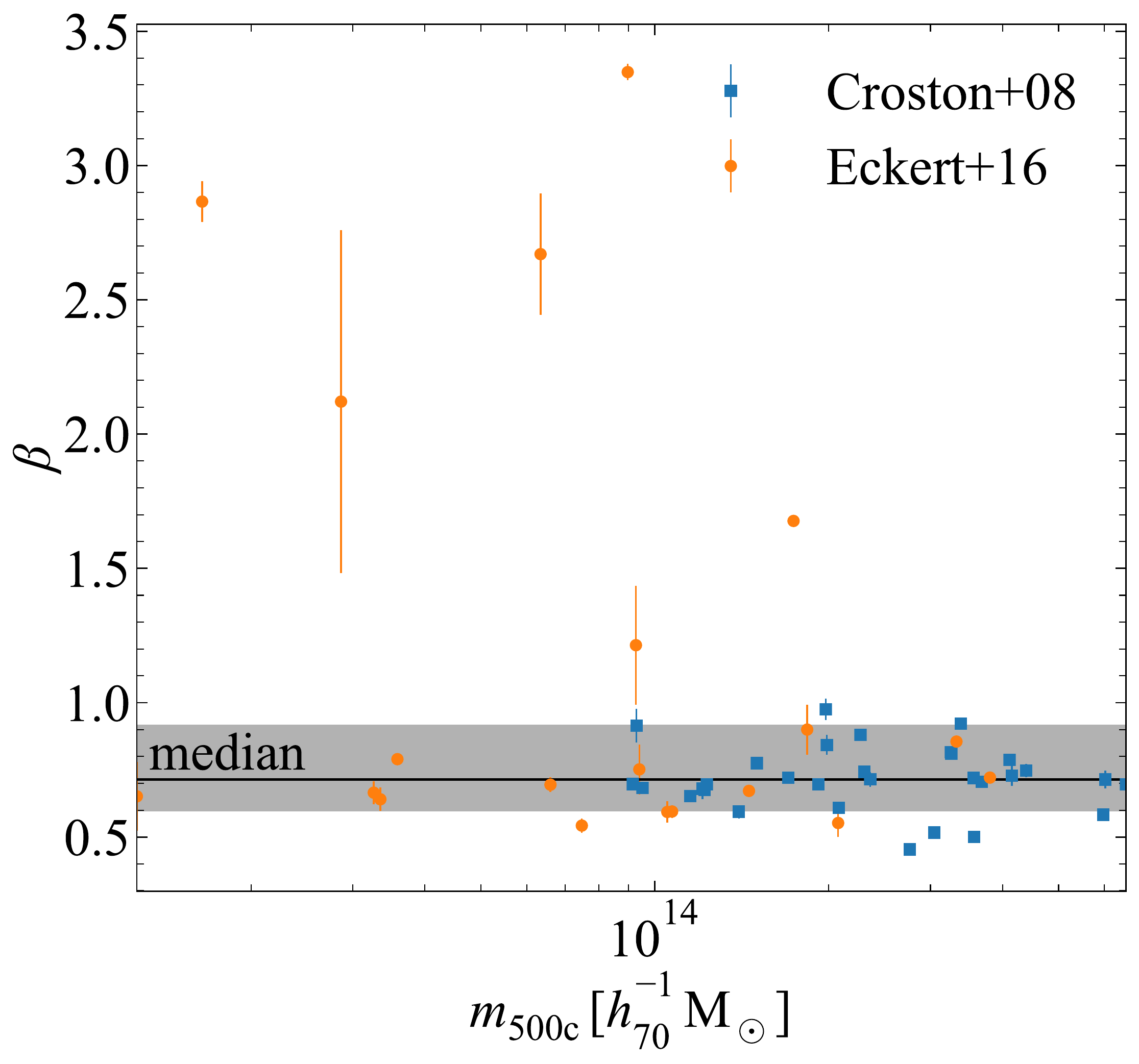}
  \caption{As Fig.~\ref{fig:obs_rc_fit} but for the slope $\beta$ of
    the beta hot gas density profile fits,
    Eq.~\ref{eq:beta_gas}.}\label{fig:obs_beta_fit}
\end{figure}
We show the core radii, $r\subscr{c}$, and slopes, $\beta$, that we
fit to our data set in Figs.~\ref{fig:obs_rc_fit} and
\ref{fig:obs_beta_fit}, respectively. There is no clear mass
dependence in the both of the fit parameters. Thus, we decided to use
the median value for both parameters for all halo masses. This
significantly simplifies the model, keeping the total number of
parameters low.

We show the hydrostatic gas fractions from our observational data in
Fig.~\ref{fig:obs_fgas}. We fit the median
\(f_{\mathrm{gas,500c}}-m_{\mathrm{500c,obs}}\) relation with a
sigmoid-like function given by
\begin{equation}
  \label{eq:fgas_sigmoid}
  f_{\mathrm{gas,500c}}(m\subscr{500c,obs}) = \frac{\Ob/\Om}{2} \left(1 +
    \tanh\left(\frac{\log_{10}(m\subscr{500c,obs}/m_\mathrm{t})}{\alpha}
    \right) \right) \, ,
\end{equation}
under the added constraint
\begin{equation}
  \label{eq:fgas_constraint}
  f_{\mathrm{gas,500c}}(m\subscr{500c,obs}) \leq
  f\subscr{b} - f\subscr{\star,500c}(m\subscr{500c,obs}) \,.
\end{equation}
The function has as free parameters the turnover mass,
\(m\subscr{t}\), and the sharpness of the turnover, \(\alpha\). We fix
the gas fraction for \(m\subscr{500c,obs}\to \infty\) to the cosmic
baryon fraction \(f\subscr{b} = \Ob/\Om \approx \num{0.166}\), which
is what we expect for deep potential wells and what we also see for
the highest-mass clusters. However, we shift down the final
$f\subscr{gas,500c}(m\subscr{500c,obs})$ relation at halo masses where
the cosmic baryon fraction would be exceeded after including the
stellar contribution. We also fix the gas fraction for \(m \to 0\) to
0 since we know that low-mass dwarfs eject their baryons easily and
are mainly dark matter dominated \citep[e.g.][]{Silk2012a,Sawala2015}.
Moreover, their virial temperatures are too low for them to contain
X-ray emitting gas. Fixing \(f\subscr{gas,500c}(m\to0) = 0\) is
probably not optimal, especially since we know that the lower mass
haloes will contain a significant warm gas
(\(\SI{e4}{\K} \lesssim T \lesssim \SI{e6}{\K}\)) component which
should increase their baryonic mass. However, since we will use our
freedom in correcting the gas fraction at \(r\subscr{h}\) by assuming
profiles outside \(r\subscr{500c,obs}\), this choice should not
significantly impact our results as we already discussed at the end of
\S~\ref{sec:hm_profiles_gas}. For our scales of interest, the shape
of the profiles of low-mass systems will not matter as much as their
total mass. Forcing the gas fraction to go to 0 for low halo masses
causes a deviation from the observations at low halo masses. However,
at low halo mass the X-ray observations will always be biased to
systems with high gas masses, since these will have the highest X-ray
luminosities.

In Fig.~\ref{fig:obs_fgas} we also show fits to the data
\(f_{\mathrm{gas,500c}}-m_{\mathrm{500c,obs}}\) relation assuming a
constant hydrostatic mass bias of
\(\frac{m\subscr{hydro}}{m\subscr{true}} = 1-b = 0.7\). In
\S~\ref{sec:results_bias} we discuss how we compute this relation
and the influence of this assumption on our results.

\section{Model density components}\label{sec:components}
\begin{figure}
  \centering
  \includegraphics[width=\columnwidth]{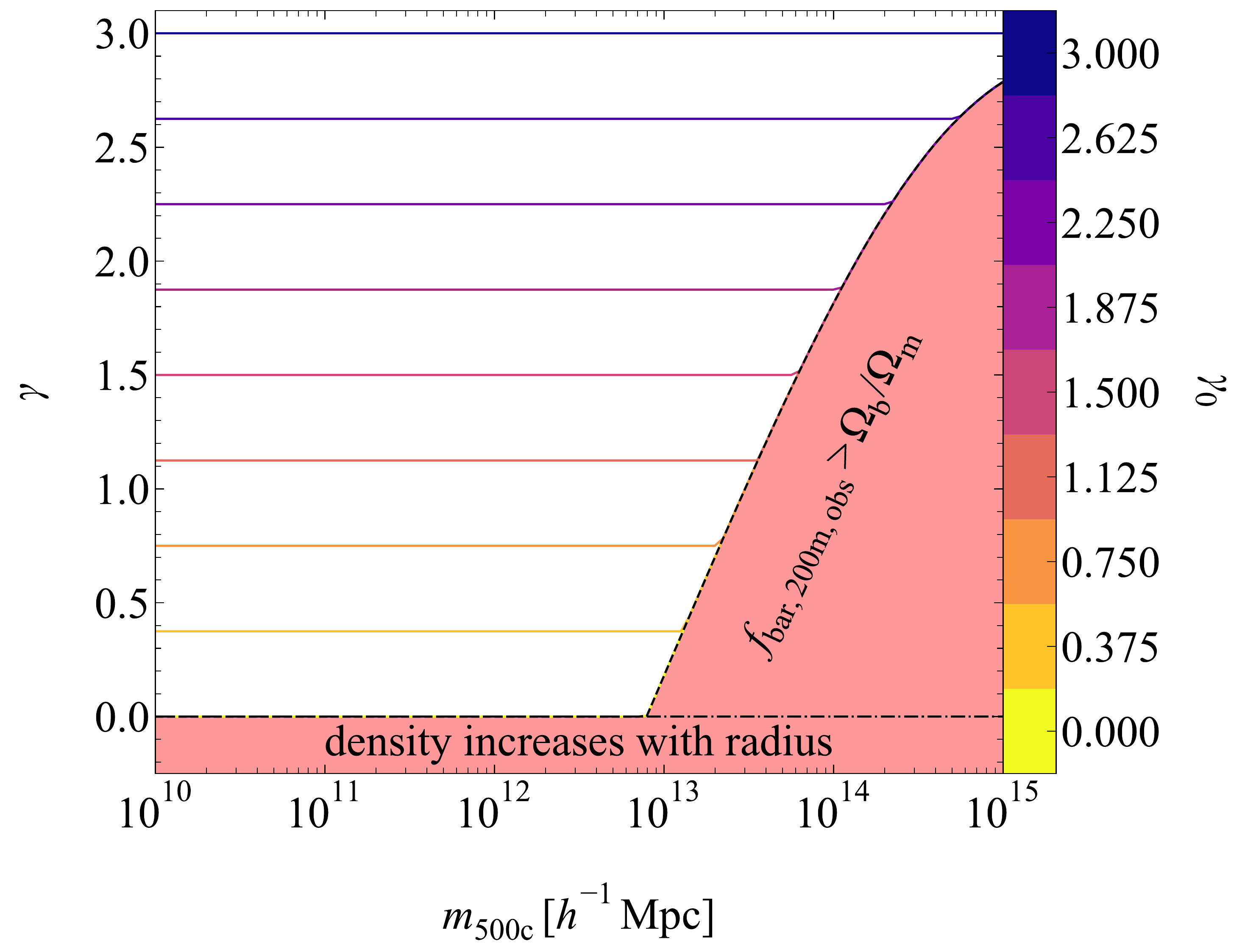}
  \caption{The allowed values for the extrapolated slope \(\gamma\) of
    the beta density profile, Eq.~\ref{eq:beta_gas}, as a function of
    halo mass \(m\subscr{500c,obs}\). We colour each line by the value
    $\gamma_0 = \gamma(m\subscr{500c,obs} \to 0)$. Since we
    extrapolate haloes to \(r\subscr{200m,obs}\), the most massive
    haloes would contain too many baryons if \(\gamma\) would be too
    small. Hence, for each halo mass, we compute the limiting
    \(\gamma\) for which the halo is baryonically closed at
    \(r\subscr{200m,obs}\). This limit is indicated by the dashed
    line. For each halo mass only slopes steeper than this limit are
    allowed.}\label{fig:gamma_0_vs_m500c}
\end{figure}
\begin{figure*}
  \centering
  \includegraphics[width=\textwidth]{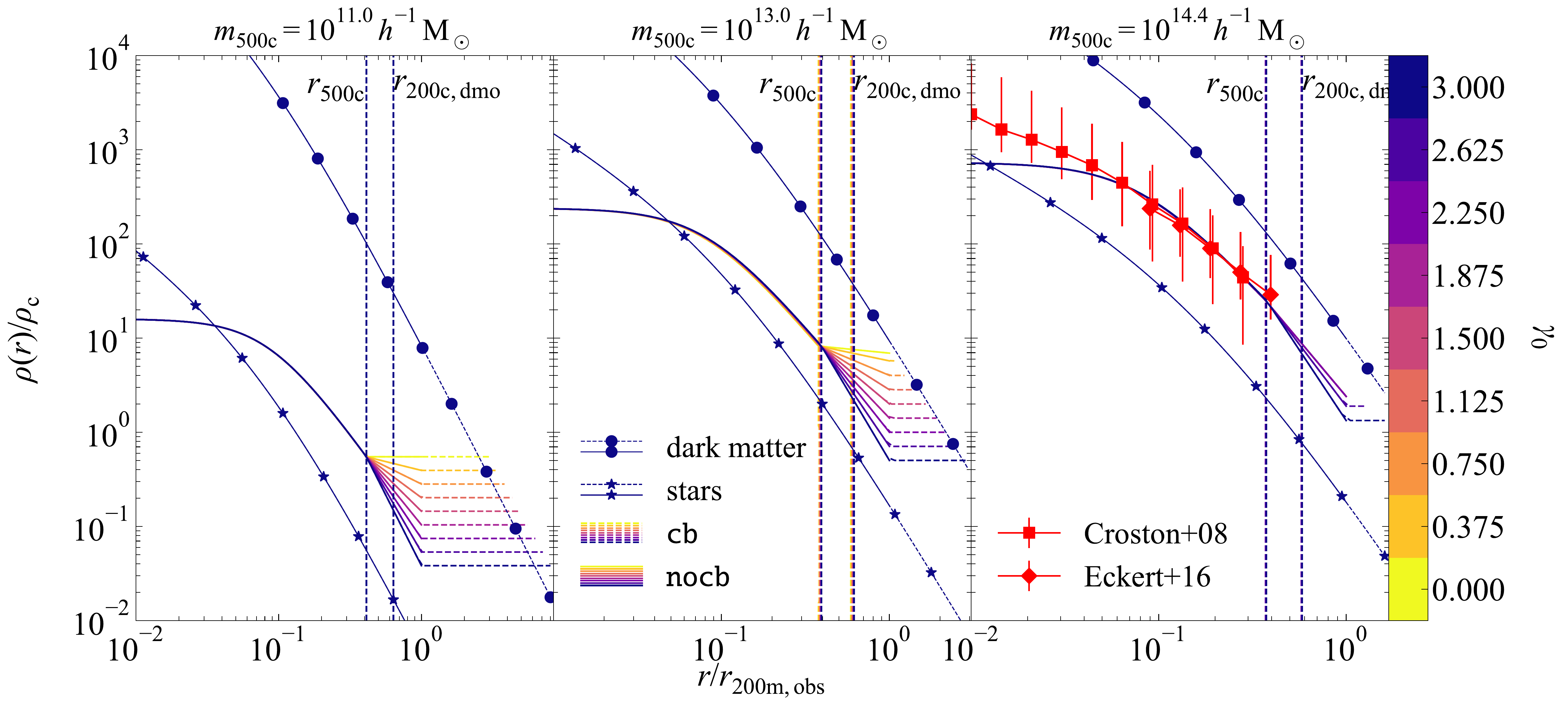}
  \caption{Model density profiles for
    $m\subscr{500c,obs} = \SI{e11}{\mh}$ (\emph{left panel}),
    $m\subscr{500c,obs} = \SI{e13}{\mh}$ (\emph{middle panel}), and
    $m\subscr{500c,obs} = \SI{e14.4}{\mh}$ (\emph{right panel}). In
    the right panel, we also show the median hot gas density profiles
    inferred from REXCESS \citep[red, connected
    squares,][]{Croston2008} and XXL-100-GC \citep[red, connected
    diamonds,][]{Eckert2016} for a sample with the same median halo
    mass. The error bars indicate the 15\supscr{th} and 85\supscr{th}
    percentile range for each radial bin. The lines are colour-coded
    by $\gamma_0 \equiv \gamma(m\subscr{500c} \to 0)$, the
    extrapolated power-law slope of the hot gas density profiles
    between $r\subscr{500c,obs}$ and $r\subscr{200m,obs}$, with lower
    values of $\gamma_0$ corresponding to flatter slopes. All profiles
    assume the best-fit beta profile up to \(r\subscr{500c,obs}\).
    Between \(r\subscr{500c,obs}\) and \(r\subscr{200m,obs}\) the
    profiles are extrapolated with a power-law slope of \(\gamma\). In
    the case of \nocb\ (solid, coloured lines), we cut off the
    profiles at the halo definition $r\subscr{200m,obs}$, at which
    point the halo baryon fraction may be smaller than $\Ob/\Om$. For
    the case \cb\ (dashed, coloured lines), we extrapolate the hot gas
    density profile with a uniform profile until the cosmic baryon
    fraction is reached. For the case \cb\ of the dark matter (dashed,
    connected circles) and stellar satellite (dashed, connected stars)
    profiles, we only show the models with $\gamma_0=3$, for the other
    values of $\gamma_0$ the maximum radius equals that of the
    corresponding hot gas profile.}
  \label{fig:obs_rho_extrapolated}
\end{figure*}
\begin{figure}
  \centering
  \includegraphics[width=\columnwidth]{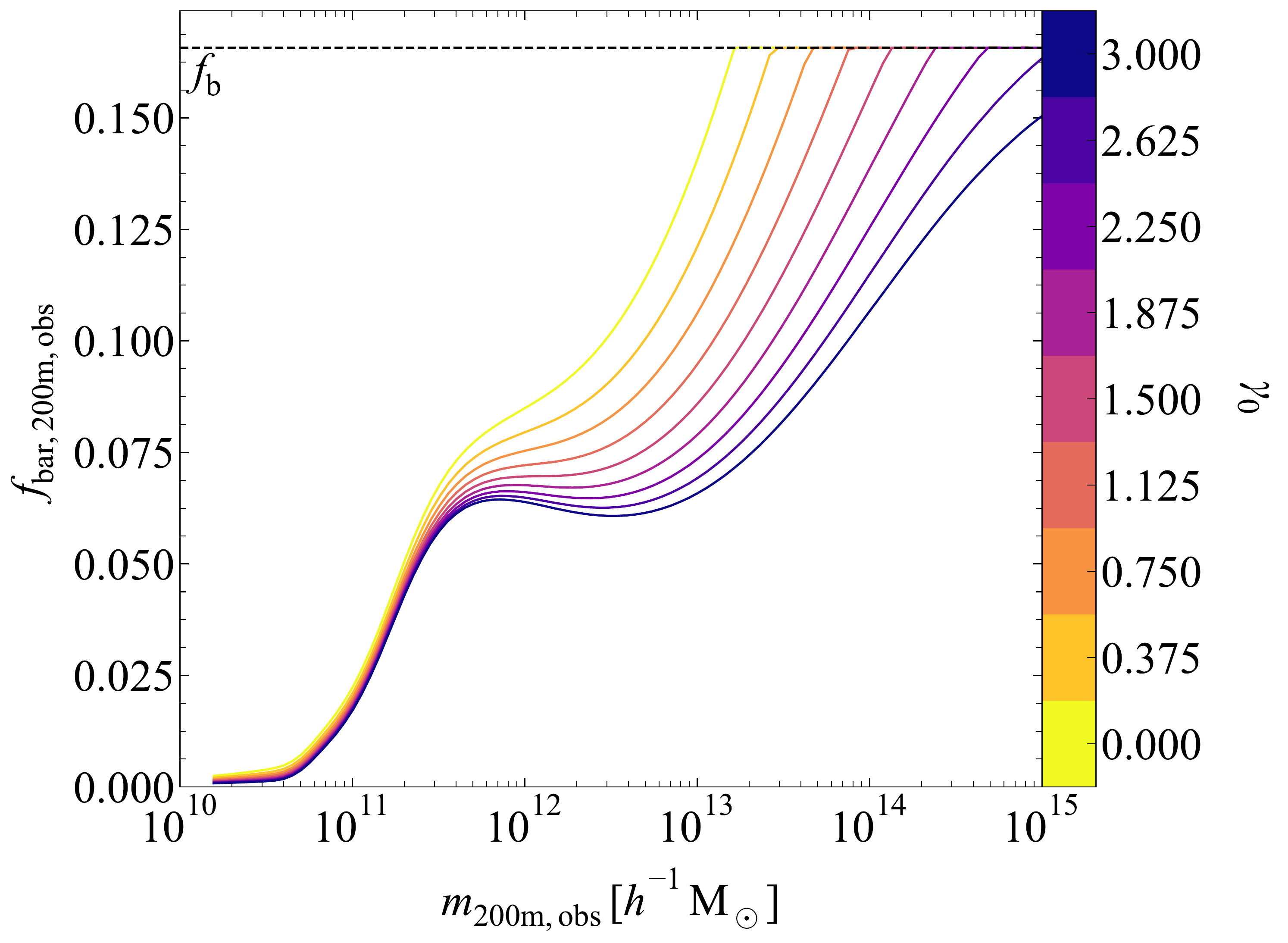}
  \caption{The halo baryon fraction at $r\subscr{200m,obs}$ as a
    function of halo mass $m\subscr{200m,obs}$. The baryon fraction
    $f\subscr{bar,200m,obs}$ is the same for both model \nocb, which
    effectively assumes that the missing halo baryons are
    redistributed far beyond $r\subscr{200m,obs}$ on linear scales,
    and model \cb, which adds the missing halo baryons in a uniform
    profile outside but near $r\subscr{200m,obs}$. The lines are
    colour-coded by $\gamma_0 \equiv \gamma(m\subscr{500c} \to 0)$,
    the extrapolated power-law slope of the hot gas density profiles
    between $r\subscr{500c,obs}$ and $r\subscr{200m,obs}$, with lower
    values of $\gamma_0$ corresponding to flatter slopes. The shape is
    set by the observed constraints on the baryon fractions at
    $r\subscr{500c,obs}$. As $\gamma_0$ decreases to 0, the halo
    baryon fractions increase. The knee at
    $m\subscr{200m,obs} \approx \SI{e12}{\mh}$ is caused by the peak
    of the stellar mass fractions. The decreased range of possible
    baryon fractions for low-mass haloes is the consequence of their
    low gas fractions and the fixed prescription for the stellar
    component.}\label{fig:fb_vs_m200m_obs}
\end{figure}
\begin{figure}
  \centering
  \includegraphics[width=\columnwidth]{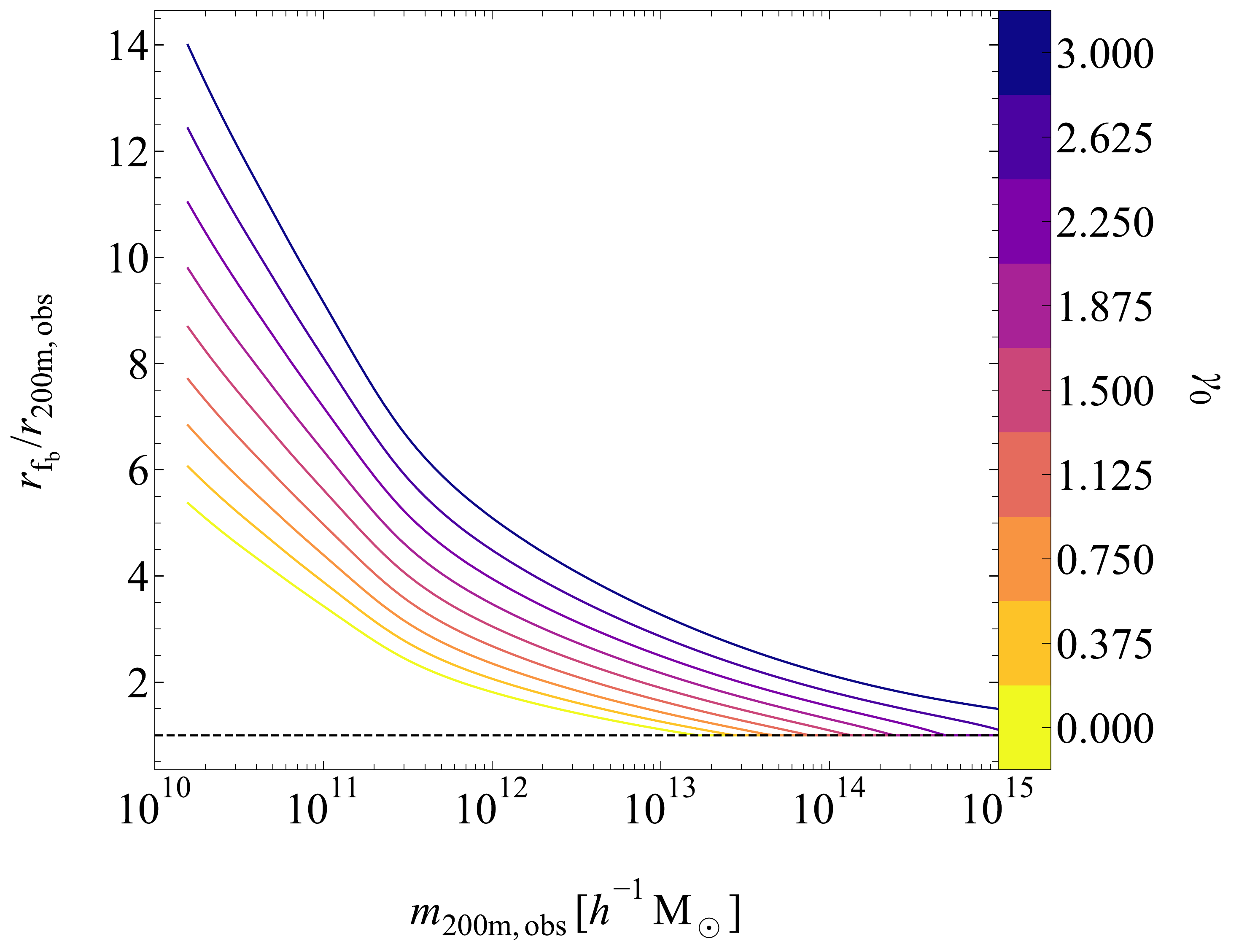}
  \caption{The radius where the cosmic baryon fraction is reached in
    units of $r\subscr{200m,obs}$ as a function of halo mass
    $m\subscr{200m,obs}$ for model \cb, which adds the missing halo
    baryons in a uniform profile outside $r\subscr{200m,obs}$. The
    lines are colour-coded by
    $\gamma_0 \equiv \gamma(m\subscr{500c} \to 0)$, the extrapolated
    power-law slope of the hot gas density profiles between
    $r\subscr{500c,obs}$ and $r\subscr{200m,obs}$, with lower values
    of $\gamma_0$ corresponding to flatter slopes. As $\gamma_0$
    decreases to 0, the cosmic baryon fraction is reached closer to
    the halo radius $r\subscr{200m,obs}$.}\label{fig:rfb_vs_m200m_obs}
\end{figure}

We determined the best-fit parameters for the beta profile,
Eq.~\ref{eq:beta_gas}, in \S~\ref{sec:obs_xray}. The only remaining
free parameter in our model is now the slope $\gamma$ of the
extrapolated profile outside $r\subscr{500c,obs}$. As we explained in
\S~\ref{sec:obs_xray}, not all values of $\gamma$ are allowed for
each halo mass $m\subscr{500c,obs}$, since the most massive haloes
contain a significant fraction of their total baryon budget inside
$r\subscr{500c,obs}$. Consequently, these haloes need steeper slopes
$\gamma$, since otherwise they would exceed the cosmic baryon fraction
before they reach the halo radius $r\subscr{200m,obs}$. We thus
determine the relation $\gamma\subscr{min}(m\subscr{500c,obs})$ that
limits the extrapolated slope such that, given the best-fit beta
profile parameters, the halo reaches exactly the cosmic baryon
fraction at $r\subscr{200m,obs}$. For each halo mass only slopes
steeper than this limiting value are allowed. We show the resulting
relation $\gamma\subscr{min}(m\subscr{500c,obs})$ in
Fig.~\ref{fig:gamma_0_vs_m500c}. We colour the curves by
$\gamma_0=\gamma(m\subscr{500c,obs} \to 0)$. Since low-mass haloes
have low baryon fractions at $r\subscr{500c,obs}$, we find that all
values of $\gamma_0$ are allowed. For the most massive haloes, only
the steepest slopes $\gamma \gtrsim 2.8$ are allowed. The handful of
observations that are able to probe clusters out to
$r\subscr{200m,obs}$ indeed find that the slope steepens in the
outskirts \citep{Ghirardini2018}.

Now we have all of the ingredients of our model at hand. We show the
resulting profiles for our different matter components for 3 halo
masses in Fig.~\ref{fig:obs_rho_extrapolated}. We show both the \nocb\
and \cb\ models, where the latter are just the former extended beyond
$r\subscr{200m,obs}$ until the cosmic baryon fraction is reached. We
colour the curves by $\gamma_0$. Given $\gamma_0$, the actual value of
the slope $\gamma$ for each halo mass can be determined by following
the tracks in Fig.~\ref{fig:gamma_0_vs_m500c} from low to high halo
masses, e.g. for the $m\subscr{500c,obs} = \SI{e15}{\mh}$ halo all
slopes $\gamma_0 \leq 2.8$ correspond to the actual slope
$\gamma=2.8$. Besides the hot gas profiles, we also show the dark
matter and stellar (satellite, since the central is modelled as a
delta function) profiles. These profiles only depend on the value
$\gamma_0$ through their maximum radius, since the halo radius
$r\subscr{h}$ is determined by how fast the cosmic baryon fraction is
reached and thus depends on $\gamma_0$.

It is clear that models with flatter slopes reach their baryon budget
at smaller radii. These models will thus capture the influence of a
compact baryon distribution on the matter power spectrum. We show the
halo baryon fraction at $r\subscr{200m,obs}$ for different values of
$\gamma_0$ in Fig.~\ref{fig:fb_vs_m200m_obs}. The main shape of the
gas fractions at $r\subscr{200m,obs}$ is set by the constraints on the
gas fractions at $r\subscr{500c,obs}$. The group-size haloes have the
largest spread in baryon fraction with changing slope $\gamma_0$. Our
model is thus able to capture a large range of different baryon
contents for haloes that all reproduce the observations at
$r\subscr{500c,obs}$. The baryon fractions rise steeply between
$\num{e11} < m\subscr{200m,obs} / (\si{\mh}) < \num{e12}$ due to the
peak in the stellar mass fraction in this halo mass range. For the
low-mass haloes, the spread in baryon fraction is smaller at
$r\subscr{200m,obs}$ because we hold the stellar component fixed in
our model and their gas fractions are low. As a result, the low-mass
systems do not differ much in the \nocb\ model. (In the \cb\ model
they will differ due to the different halo radii $r\subscr{h}$ where
the cosmic baryon fraction is reached.) For the slope $\gamma$ between
$\numrange{0}{3}$ we will have $\approx \SIrange{20}{50}{\percent}$ of
the total baryons in the Universe outside haloes in the \nocb\ model.

We have checked that the density profiles with varying $\gamma_0$ for
for haloes with
$\num{e14} < m\subscr{500c,obs} / \si{\mh} < \num{e15}$ only cause a
maximum deviation of $\approx \pm \SI{5}{\percent}$ in the surface
brightness profiles for projected radii $R < r\subscr{500c,obs}$
compared to the fiducial model with $\gamma_0 = 3\beta$. This
variation is within the error on the surface brightness counts and the
density profiles with varying $\gamma_0$ are thus indistinguishable
from the fiducial model in the investigated mass range. For haloes
with $m\subscr{500c,obs} \leq \SI{e14}{\mh}$, the deviations increase
for lower values of $\gamma_0$, reaching $\SI{10}{\percent}$ for
$\gamma_0=1.5$ and $m\subscr{500c,obs} = \SI{e13}{\mh}$, but the
observed hot gas density profiles at these halo masses also show a
larger scatter.

We also have the \cb\ model where we force all haloes to include all
of the missing baryons in their outskirts. In
Fig.~\ref{fig:rfb_vs_m200m_obs} we show how extended the baryon
distribution needs to be in the \cb\ case as a function of the slope
$\gamma_0$. The variations in the power-law slope paired with the \cb\
and \nocb\ models allow us to investigate the influence on the matter
power spectrum of a wide range of possible baryon distributions that
all reproduce the available X-ray observations for clusters with
$m\subscr{500c,obs} \geq \SI{e14}{\mh}$.

\section{Results}\label{sec:results}
In this section we show the results and predictions of our model for
the matter power spectrum and we discuss their implications for future
observational constraints. First, we show the influence of assuming
different distributions for the unobserved hot gas in
\S~\ref{sec:results_outer}. We show the influence of correcting
observed halo masses to the dark matter only equivalent halo masses in
order to obtain the correct halo abundances in
\S~\ref{sec:results_hmf}. In \S~\ref{sec:results_masses}, we show
which halo masses dominate the power spectrum for which wavenumbers.
Finally, we show the influence of varying the best-fit observed
profile parameters in \S~\ref{sec:results_prms} and we investigate
the effects of a hydrostatic bias in the halo mass determination in
\S~\ref{sec:results_bias}.

\subsection{Influence of the unobserved baryon
  distribution}\label{sec:results_outer}
\begin{figure*}
  \centering
  \includegraphics[width=\textwidth]{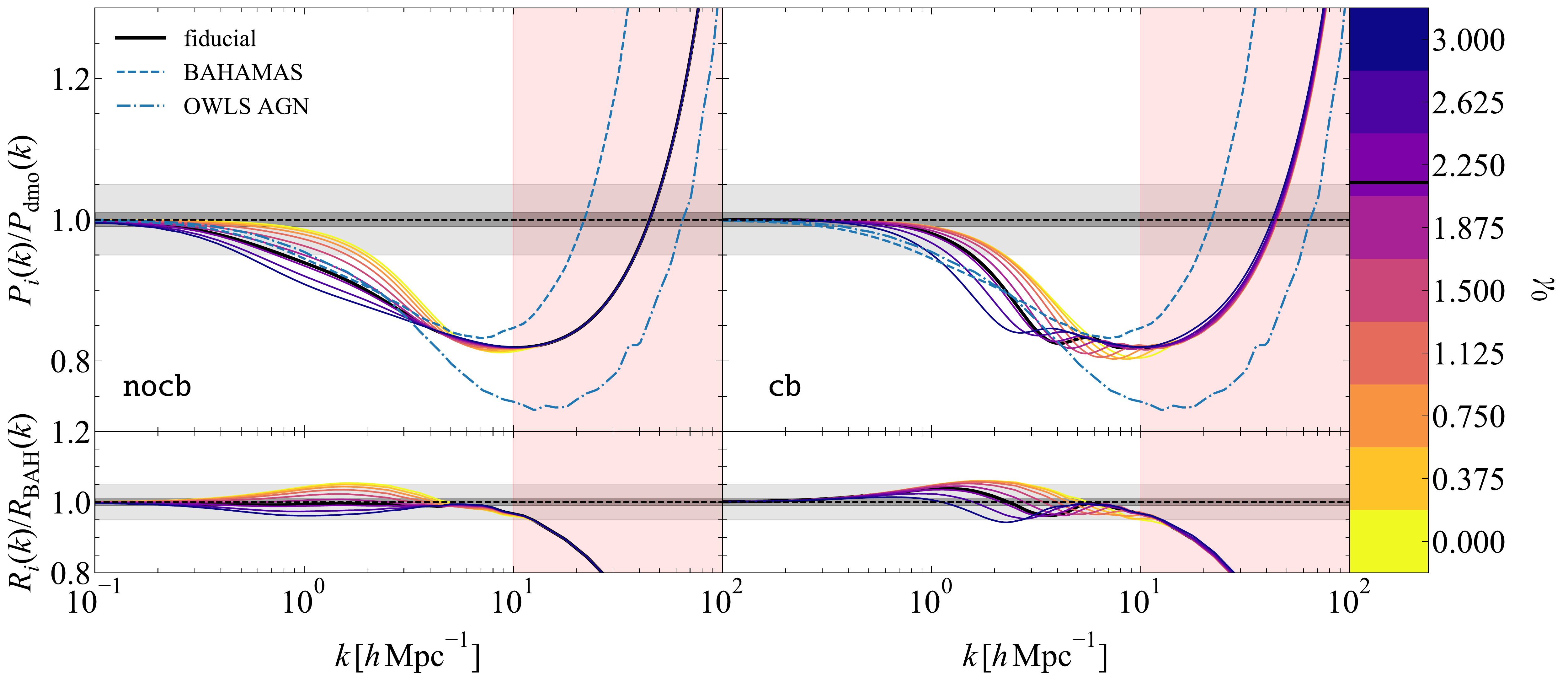}
  \caption{\emph{Top row} The ratio of our halo model power spectra
    with baryons to the corresponding dark matter only prediction. The
    dark and light gray bands indicate the $\SI{1}{\percent}$ and
    $\SI{5}{\percent}$ intervals. The left-hand panel shows model
    \nocb, which effectively assumes that the baryons missing from
    haloes are redistributed far beyond $r\subscr{200m,obs}$ on linear
    scales. The right-hand panel shows model \cb, which adds the
    missing halo baryons in a uniform profile outside but near
    $r\subscr{200m,obs}$. The red, lightly-shaded region for
    $k > \SI{10}{\impch}$ indicates the scales where our model is is
    not a good indicator of the uncertainty because the stellar
    component is not varied. We indicate our fiducial model, which
    simply extrapolates the best-fit beta profile to the hot gas
    density profiles of clusters, with a thick black line. The lines
    are colour-coded by
    $\gamma_0 \equiv \gamma(m\subscr{500c} \to 0)$, the extrapolated
    power-law slope of the hot gas density profiles between
    $r\subscr{500c,obs}$ and $r\subscr{200m,obs}$, with lower values
    of $\gamma_0$ corresponding to flatter slopes. The clear
    difference between the \nocb\ and \cb\ models on large scales
    indicates that it is very important to know where the missing halo
    baryons end up. Placing the missing halo baryons in the vicinity
    of the haloes increases the power on large scales significantly
    for fixed $\gamma_0$. Our model is flexible enough, especially in
    the \nocb\ case, to encompass the behaviour of both the \bah\
    (blue, dashed line) and \owls\ (blue, dash-dotted line)
    simulations on large scales. \emph{Bottom row} The ratio between
    the matter power spectra responses to baryons predicted by our
    models and the \bah\ simulation. Both our fiducial models are
    within \(\approx \SI{5}{\percent}\) of \bah\ for
    \(k < \SI{10}{\impch}\).}\label{fig:power_ratio}
\end{figure*}
\begin{figure*}
  \centering
  \includegraphics[width=\textwidth]{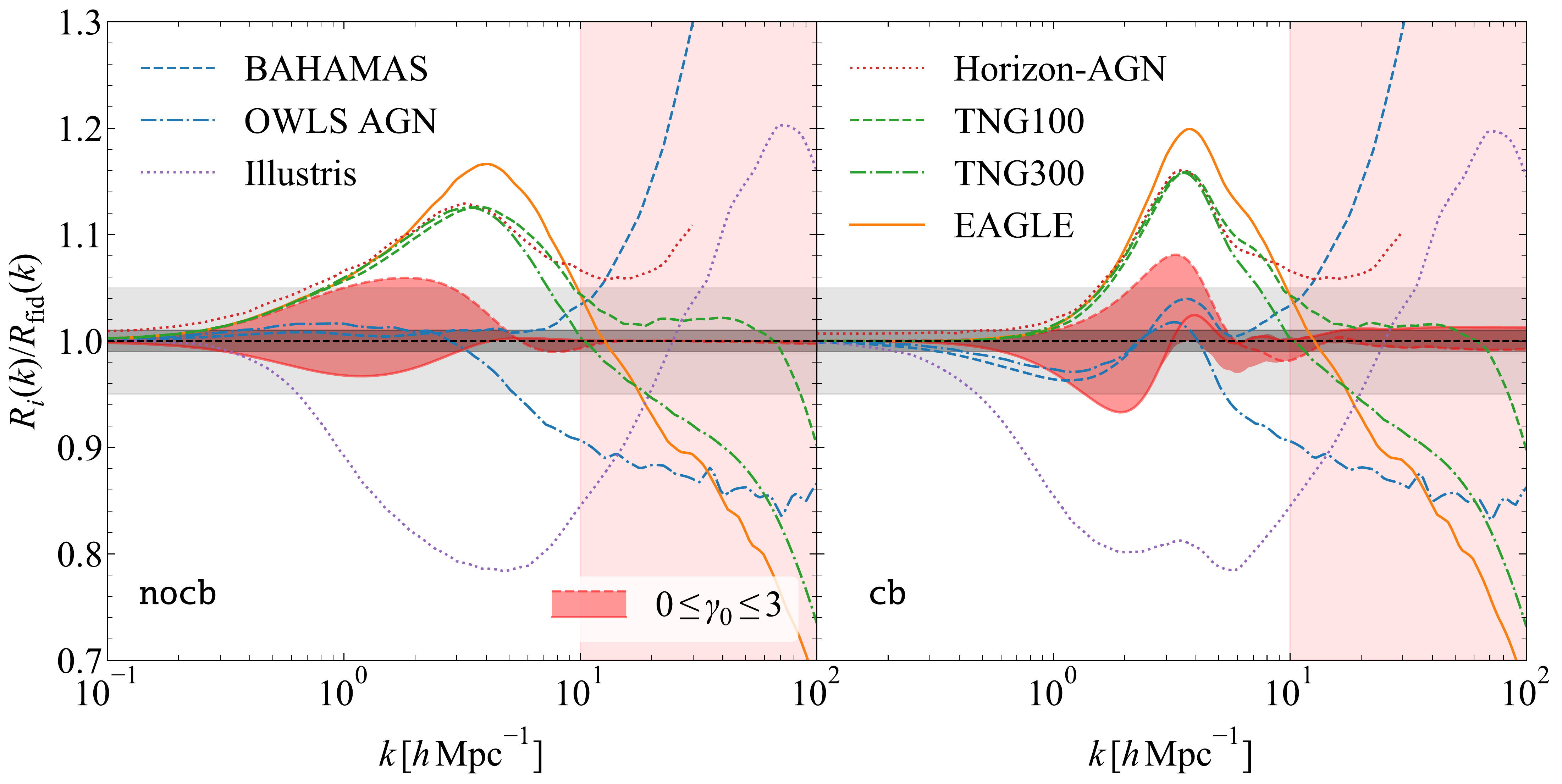}
  \caption{The ratio between the power spectrum response to baryons
    predicted by hydrodynamical simulations from the literature and
    our fiducial halo model, which simply extrapolates the best-fit
    beta profile to the hot gas density profiles of clusters. The dark
    and light gray bands indicate the $\SI{1}{\percent}$ and
    $\SI{5}{\percent}$ intervals. The left-hand panel shows model
    \nocb, which effectively assumes that the baryons missing from
    haloes are redistributed far beyond $r\subscr{200m,obs}$ on linear
    scales. The right-hand panel shows model \cb, which adds the
    missing halo baryons in a uniform profile outside but near
    $r\subscr{200m,obs}$. The red, lightly-shaded region for
    $k > \SI{10}{\impch}$ indicates the scales where our model is is
    not a good indicator of the uncertainty because the stellar
    component is not varied. The shaded red region shows the spread in
    our models for all values of $\gamma_0$, i.e. the extrapolated
    power-law slope of the hot gas density profiles between
    $r\subscr{500c,obs}$ and $r\subscr{200m,obs}$, with red lines
    indicating $\gamma_0=0$ (red, dashed line) and $\gamma_0=3$ (red,
    solid line). The blue lines indicate simulations that reproduce
    the hot gas fractions of clusters, i.e. \bah\ \citep[blue, dashed
    line,][]{McCarthy2017} and \owls\ AGN \citep[blue, dash-dotted
    line,][]{VanDaalen2011}. We show the higher-resolution but
    smaller-volume simulations which predict too large cluster gas
    fractions, i.e. EAGLE \citep[orange, solid line,][]{Hellwing2016},
    IllustrisTNG-100 and 300 \citep[green, dashed and dash-dotted
    lines, respectively,][]{Springel2017}, Horizon-AGN \citep[red,
    dotted line,][]{Chisari2018}. We also show the original Illustris
    result \citep[purple, dotted line,][]{Vogelsberger2014a}, which
    underpredicts cluster gas fractions. Our empirical model
    encompasses the simulations that reproduce the cluster hot gas
    fractions on all scales $k \lesssim \SI{5}{\impch}$, but the other
    simulations fall outside of the allowed
    range.}\label{fig:power_ratio_sims}
\end{figure*}
\begin{figure}
  \centering
  \includegraphics[width=\columnwidth]{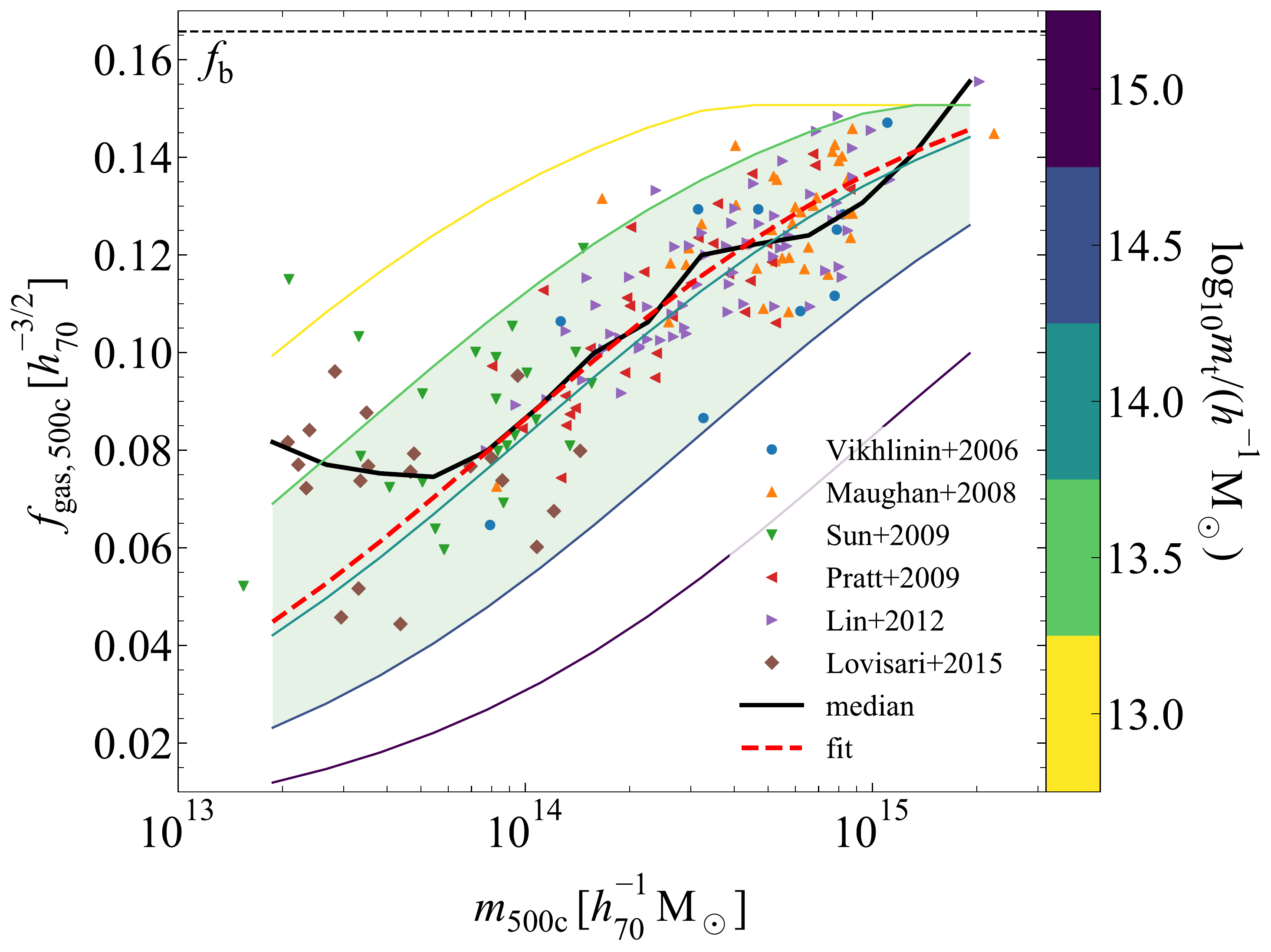}
  \caption{The X-ray hydrostatic gas fractions as a function of halo
    mass, as in Fig.~\ref{fig:obs_fgas}. The curves show the
    sigmoid-like fit from Eq.~\eqref{eq:fgas_sigmoid} with the
    best-fit value for $\alpha=1.35$, coloured by the value
    $\log_{10}m\subscr{t}/(\si{\mh}) \in \{13,13.5,14,14.5,15\}$ (the
    best-fit value is 13.94). The shaded green region indicates the
    area that is broadly in agreement with
    observations.}\label{fig:fgas_vs_mt}
\end{figure}
\begin{figure}
  \centering
  \includegraphics[width=\columnwidth]{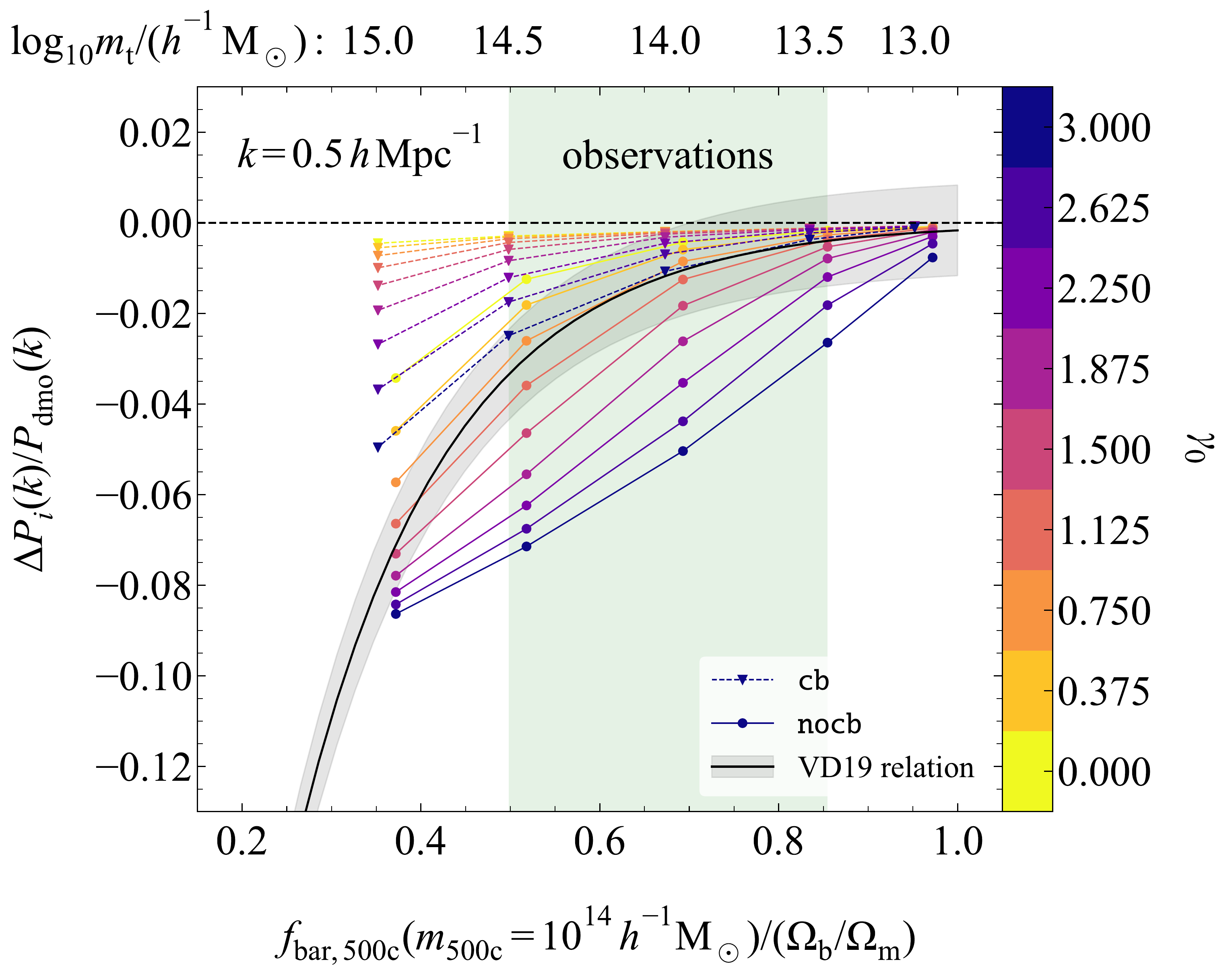}
  \caption{The power suppression due to the inclusion of baryons at
    the fixed scale $k = \SI{0.5}{\impch}$ as a function of the baryon
    fraction of haloes with $m\subscr{500c,obs}=\SI{e14}{\mh}$. The
    shaded green region indicates the gas fractions that broadly agree
    with observations. The \cb\ (dashed, connected triangles) and
    \nocb\ (connected circles) models are coloured by $\gamma_0$, i.e.
    the value of the extrapolated power-law slope of the hot gas
    density profiles between $r\subscr{500c,obs}$ and
    $r\subscr{200m,obs}$. We show the relation found by
    \citet{VanDaalen2019} for hydrodynamical simulations and its
    $\pm \SI{1}{\percent}$ variation (black line with grey, shaded
    region). We indicate the value of
    $\log_{10} m\subscr{t}/(\si{\mh})$ in Eq.~\eqref{eq:fgas_sigmoid}
    along the top x-axis. Both our model and VD19 predict a positive
    correlation between the power suppression at fixed scale and the
    halo baryon fraction at fixed halo mass. However, it is clear that
    our model allows for a larger range in possible power suppression
    at fixed halo baryon fraction than is found in the
    simulations.}\label{fig:power_vs_fbar}
\end{figure}
In this section, we will investigate the influence of the distribution
of the unobserved baryons inside and outside haloes on the matter
power spectrum. Since we currently have only a very tenuous grasp of
the whereabouts of the missing baryons, it is important to explore how
their possible distribution impacts the matter power spectrum.

As stated in \S~\ref{sec:hm_profiles_gas}, our model is characterized
by the extrapolated power-law slope $-\gamma$ for the hot gas density
profile and by whether we assume the missing halo baryons to reside in
the vicinity of the halo (model \cb) or not (model \nocb). As
explained in \S~\ref{sec:hm_profiles_gas}, these two types of models
only differ in $\hat{\rho}(k|m)$ due to the inclusion of more mass
outside the traditional halo definition of $r\subscr{200m,obs}$ in the
\cb\ case (see Figs.~\ref{fig:obs_rho_extrapolated},
\ref{fig:rfb_vs_m200m_obs}). When discussing our model predictions for
the power spectrum, we consider the range
$\SI{0.1}{\impch} \leq k \leq \SI{5}{\impch}$ to be the vital regime
since future surveys will gain their optimal signal-to-noise for
$k \approx \SI{1}{\impch}$ \citep{Amendola2018}.

We show the response of the matter power spectrum to baryons for the
\nocb\ and \cb\ models in, respectively, the top-left and top-right
panels of Fig.~\ref{fig:power_ratio}. The lines are coloured by the
assumed value of $\gamma_0$. We indicate our fiducial model, which
extrapolates the best-fit $\beta=0.71^{+0.20}_{-0.12}$, i.e.
$\gamma_0 = 3\beta = 2.14$, from the X-ray observations, with the
thick, black line. All models show a suppression of power on large
scales with respect to the DMO prediction. All of our models have an
upturn in the response for $k \gtrsim \SI{10}{\impch}$ and and
enhancement of power for $k \gtrsim \SI{50}{\impch}$ due to the
stellar component. This upturn is not present in other halo model
approaches that only modify the dark matter profiles
\citep[e.g.][]{Smith2003,Mead2015}. We shade the region
$k > \SI{10}{\impch}$ in red because the range in responses of our
model does not span the range allowed by observations there. On the
contrary, on these small scales all of our models behave the same,
since the hot gas is completely determined by the best-fit beta
profile to the X-ray observations, and the stellar component is held
fixed.

The total amount of power suppression at large scales depends
sensitively on the halo baryon fractions, since models with the
highest values of $\gamma_0$ also have the lowest baryon fractions
$f\subscr{bar,200m,obs}$ at all halo masses (see
Fig.~\ref{fig:fb_vs_m200m_obs}). Our results confirm the predictions
from hydrodynamical simulations, which have shown similar trends
\citep{VanDaalen2011, VanDaalen2019, Hellwing2016, McCarthy2017,
  Springel2017, Chisari2018}. However, our results do not rely on the
uncertain assumptions associated with subgrid models for feedback
processes. Our phenomenological model simply requires that we
reproduce the density profiles of clusters without any assumptions
about the underlying physics that resulted in the profiles.

The \nocb\ model, shown in the left-hand panel of
Fig.~\ref{fig:power_ratio}, results in a larger spread of possible
responses because the final total halo mass is not fixed to account
for all the baryons as in \cb. The \nocb\ models with the steepest
extrapolated density profiles, i.e. the highest values for $\gamma_0$,
function as upper limits on the response, since the missing halo
baryons are in reality likely to reside in the vicinity of the haloes
and because low-mass haloes likely contain more gas than predicted by
our extrapolated relation. However, this gas may not be well described
by our beta profile assumption derived from the hot gas properties of
clusters. On the other hand, the \cb\ models with flatter slopes
(lower values for $\gamma_0$), shown in the right-hand panel of
Fig.~\ref{fig:power_ratio}, function as lower limits on the response
of the power spectrum to baryons, since it is likely that a
significant fraction of the baryons does not reside inside haloes but
rather in the diffuse, warm-hot, intergalactic medium \citep[WHIM, as
has been predicted by simulations and recently inferred from
observations, see e.g.][]{Cen1999,Dave2001,Nicastro2018}. Hence, we
find that the (minimum, fiducial, maximum) value of the minimum
wavenumber for which the baryonic effect reaches $\SI{1}{\percent}$ is
($0.2$, $0.3$, $0.9$) $\si{\impch}$ in the \nocb\ models and ($0.5$,
$0.8$, $1$) $\si{\impch}$ in the \cb\ models. The $\SI{5}{\percent}$
threshold is reached for ($0.5$, $0.8$, $2$) $\si{\impch}$ and ($1$,
$1.4$, $2$) $\si{\impch}$, respectively, for the \nocb\ and \cb\
models.

We indicate the results from the \bah\ simulation run
$\mathtt{AGN\_TUNED\_nu0\_L400N1024\_WMAP9}$, which has been shown to
reproduce a plethora of observations for massive systems
\citep{McCarthy2017,Jakobs2017}, and the result for the \owls\ AGN
simulation \citep{Schaye2010,VanDaalen2011} which has been widely used
as a reference model in weak lensing analyses and is also consistent
with the observed cluster gas fractions \citep{McCarthy2010}. We show
the ratio between our models and the \bah\ prediction of the power
spectrum response to the presence of baryons in the bottom row of
Fig.~\ref{fig:power_ratio}. Our models encompass both the \bah\ and
\owls\ predictions for $k \lesssim \SI{5}{\impch}$, which is the range
of interest here. In the \cb\ case, our models all predict less power
suppression than the simulations on large scales
$k \lesssim \SI{1}{\impch}$, which is most likely due to the fact that
in the simulations there are actually baryons in the cosmic web that
should not be accounted for by haloes, thus suggesting that models
\nocb\ may be more realistic. However, since there are no
observational constraints on the location of the missing halo baryons,
we cannot exclude the models \cb. We stress that we did not fit our
model to reproduce these simulations. The overall similarity is caused
by the simulations reproducing the measured X-ray hot gas fractions
that we fit our model to.

In Fig.~\ref{fig:power_ratio_sims}, we compare predictions for the
power spectrum response to baryons from a large set of
higher-resolution, but smaller-volume, cosmological simulations to the
prediction of our fiducial model. We compare the EAGLE
\citep{Schaye2015,Hellwing2016}, IllustrisTNG \citep{Springel2017},
Horizon-AGN \citep{Chisari2018}, and Illustris
\citep{Vogelsberger2014} simulations. We can see that in all of these
simulations, except for Illustris, which is known to have AGN feedback
that is too violent on group and cluster scales
\citep{Weinberger2017}, the baryonic suppression becomes significant
only at much smaller scales than in \owls, \bah\ and our own model.
From the halo model it is clear that the total baryon content of
haloes, and thus the cluster gas fractions, are the dominant cause of
baryonic power suppression on large scales
$k \lesssim \SI{1}{\impch}$, since $\hat{\rho}(k|m) \to m$ there.
Indeed, \citet{VanDaalen2019} explicitly demonstrated the link between
cluster gas fractions and power suppression on large scales for a
large set of hydrodynamical simulations including these. Since \bah\
and \owls\ AGN reproduce the cluster hot gas fractions, they predict
the same large-scale behaviour for the power spectrum response to
baryons. However, the other small-volume, high-resolution simulations
overpredict the baryon content of groups and clusters as was shown for
EAGLE, IllustrisTNG, and Horizon-AGN by, respectively,
\citet{Barnes2017}, \citet{Barnes2018}, and \citet{Chisari2018}. We
thus stress the importance of using simulations that are calibrated
towards the relevant observations when training or comparing models
aimed at predicting the matter power spectrum.

The small-scale behaviour of the power spectrum response to baryons is
very sensitive to the stellar density profiles and as a result we see
a large variation between the different simulation predictions in
Fig.~\ref{fig:power_ratio_sims}. As is shown by \citet{VanDaalen2019},
the small-scale power turnover in the simulations depends strongly on
the resolution and subgrid physics of the simulation. We mentioned
earlier that our model is fixed at these scales by the best-fit beta
profiles to the X-ray observations and the fixed stellar component.

Recently, \citet{VanDaalen2019} analyzed 92 hydrodynamical
simulations, including all the ones shown in
Fig.~\ref{fig:power_ratio_sims}, and showed that there is a strong
correlation between the total power suppression at a fixed scale
$k \lesssim \SI{1}{\impch}$ and the baryon fraction at
$r\subscr{500c}$ of haloes with $m\subscr{500c} = \SI{e14}{\mh}$. We
investigate the same relation with our model. We show the different
relations that we assume for the gas fraction
$f\subscr{gas,500c}(m\subscr{500c,obs})$ in Fig.~\ref{fig:fgas_vs_mt}.
For these relations we assume the best-fit value $\alpha=1.35$ from
our fit to the observed gas fractions in Eq.~\eqref{eq:fgas_sigmoid},
but we vary the turnover mass from its best-fit value of
$\log_{10}m\subscr{t}/(\si{\mh})=13.94$. Thus, we can capture a large
range of possible gas fractions at $r\subscr{500c,obs}$, allowing us
to encompass both the observed and the simulated gas fractions of
$m\subscr{500c,obs}=\SI{e14}{\mh}$ haloes. For all these relations we
then compute the power spectrum response due to the inclusion of
baryons at the fixed scale $k=\SI{0.5}{\impch}$. We show the power
suppression at this scale as a function of the halo baryon fraction in
$m\subscr{500c,obs}=\SI{e14}{\mh}$ haloes in
Fig.~\ref{fig:power_vs_fbar}. Similarly to \citet{VanDaalen2019}, we
find that higher baryon fractions at fixed halo mass result in smaller
power suppression at fixed scale. In the \nocb\ (\cb) case, the model
with $\gamma_0=1.125$ ($\gamma_0=3$) most closely tracks the
prediction from the hydrodynamical simulations. However, since our
model has complete freedom for the gas density profile in the halo
outskirts, the range of possible power suppression is much larger than
that found in the simulations analyzed by \citet{VanDaalen2019}. The
matter distribution in simulations is constrained by the subgrid
physics that is assumed. Hence, relying only on simulation predictions
might result in an overly constrained and model-dependent parameter
space, since other subgrid recipes might result in differences in the
matter distribution at large scales.

We conclude that the total baryon fraction of massive haloes is of
crucial importance to the baryonic suppression of the power spectrum.
Our model and hydrodynamical simulations that reproduce the cluster
gas fractions are in general agreement about the total amount of
suppression at scales $k \lesssim \SI{5}{\impch}$, with the exact
amplitude depending on the details of the missing baryon distribution
and varying by $\approx \pm \SI{5}{\percent}$ around our fiducial
model. Observations of the total baryonic mass for a large sample of
groups and clusters would provide a powerful constraint on the effects
of baryons on the matter power spectrum, provided we are able to
reliably measure the cluster masses. Cluster gas masses can be
determined with X-ray observations and their outskirts can be probed
with SZ measurements. Groups are subject to a significant Malmquist
bias in the X-ray regime and SZ measurements from large surveys like
Planck \citep{PlanckXXVII2016}, the Atacama Cosmology Telescope
\citep[ACTPol,][]{Hilton2017}, and the South Pole Telescope
\citep[SPT,][]{Bleem2015} generally do not reach a high enough
Signal-to-Noise ratio (SNR) to reliably measure the hot gas properties
of group-mass haloes. Constraining the total baryon fraction of these
haloes is thus challenging. However, progress could be made by
adopting cross-correlation approaches between SZ maps and large
redshift surveys as in \citet{Lim2018}. Finally, accurately
determining the baryon fraction relies on accurate halo mass
determinations for the observed systems. Halo masses can be determined
from scaling relations between observed properties (e.g. the hot gas
mass, the X-ray temperature, or the X-ray luminosity) and the total
halo mass. However, these relations need to be calibrated to a direct
measurement of the halo mass through e.g. a weak lensing total mass
profile. We will investigate the influence of a hydrostatic bias in
the halo mass determination in \S~\ref{sec:results_bias}.

\subsection{Influence of halo mass correction due to baryonic
  processes}\label{sec:results_hmf}
\begin{figure}
  \centering
  \includegraphics[width=\columnwidth]{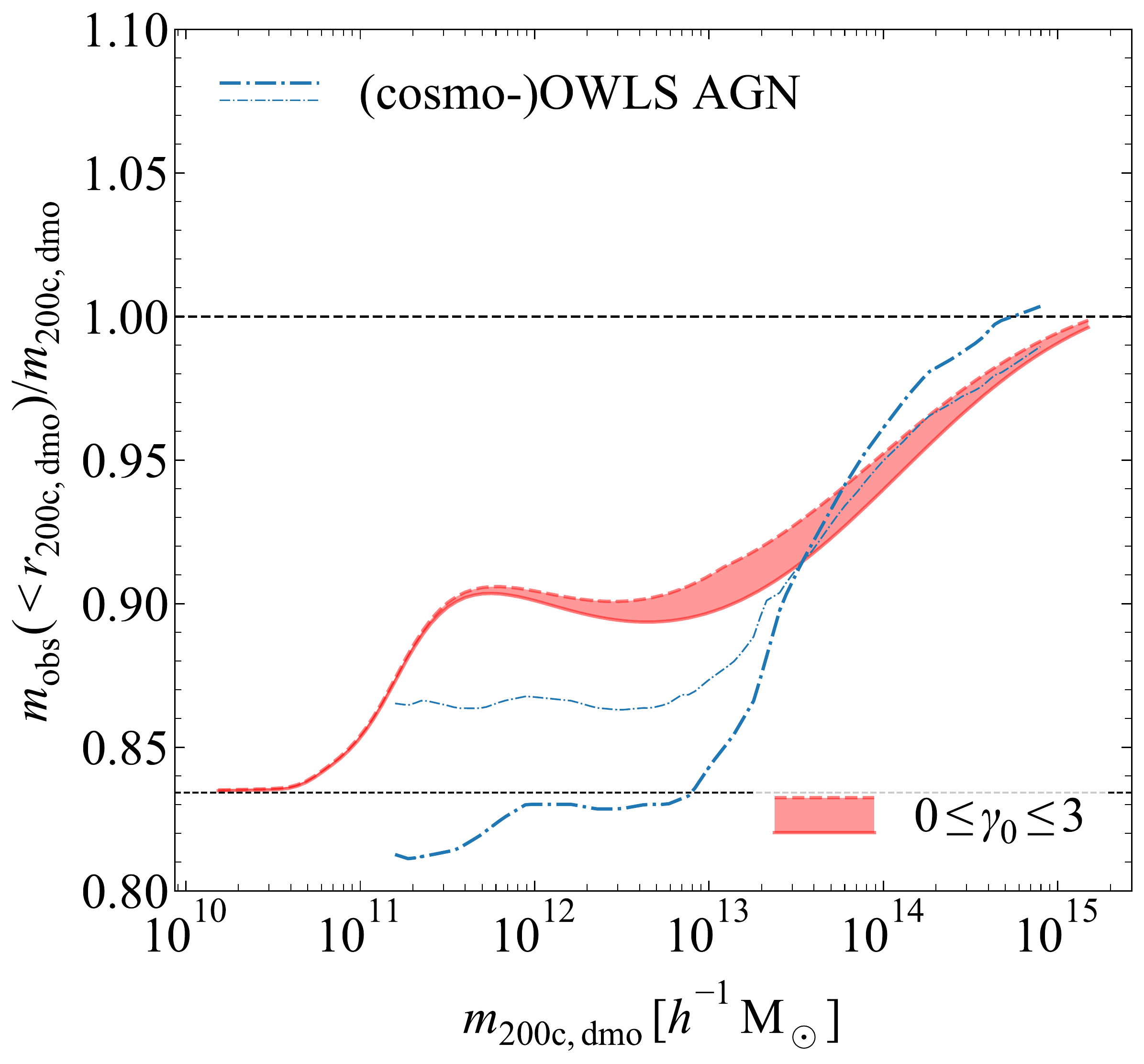}
  \caption{The ratio of the enclosed observed halo mass to the dark
    matter only equivalent mass at the fixed radius
    $r\subscr{200c,dmo}$ as a function of $m\subscr{200c,dmo}$. The
    shaded red region shows the spread in our models for all values of
    $\gamma_0$, i.e. the extrapolated power-law slope of the hot gas
    density profiles between $r\subscr{500c,obs}$ and
    $r\subscr{200m,obs}$, with red lines indicating $\gamma_0=0$ (red,
    dashed line) and $\gamma_0=3$ (red, solid line). The thin, black,
    dotted line indicates the ratio $1 - f\subscr{b}$ that our model
    converges to when the halo baryon fraction reaches 0. The mass
    ratios at fixed radius $r\subscr{200c,dmo}$ converge towards high
    halo masses since not all values of $\gamma$ are allowed for
    massive haloes. For low masses, the ratios converge because the
    stellar component is held fixed and the gas fractions are low. The
    thick, blue, dash-dotted line shows the same relation at fixed
    radius $r\subscr{200c,dmo}$ in the (cosmo-)\owls\ AGN simulation
    \citep{Velliscig2014}. The thin, blue, dash-dotted line shows the
    simulation relation corrected for changes in the dark matter mass
    profiles at $r\subscr{200c,dmo}$ with respect to the DMO
    equivalent haloes, since our model assumes that baryons do not
    affect the dark matter profile. The remaining difference in the
    mass ratio is due to differing baryon fractions between our model
    and the simulations.}\label{fig:mobs_r200c_dmo_m200c_dmo_ratio}
\end{figure}
\begin{figure*}
  \centering
  \includegraphics[width=\textwidth]{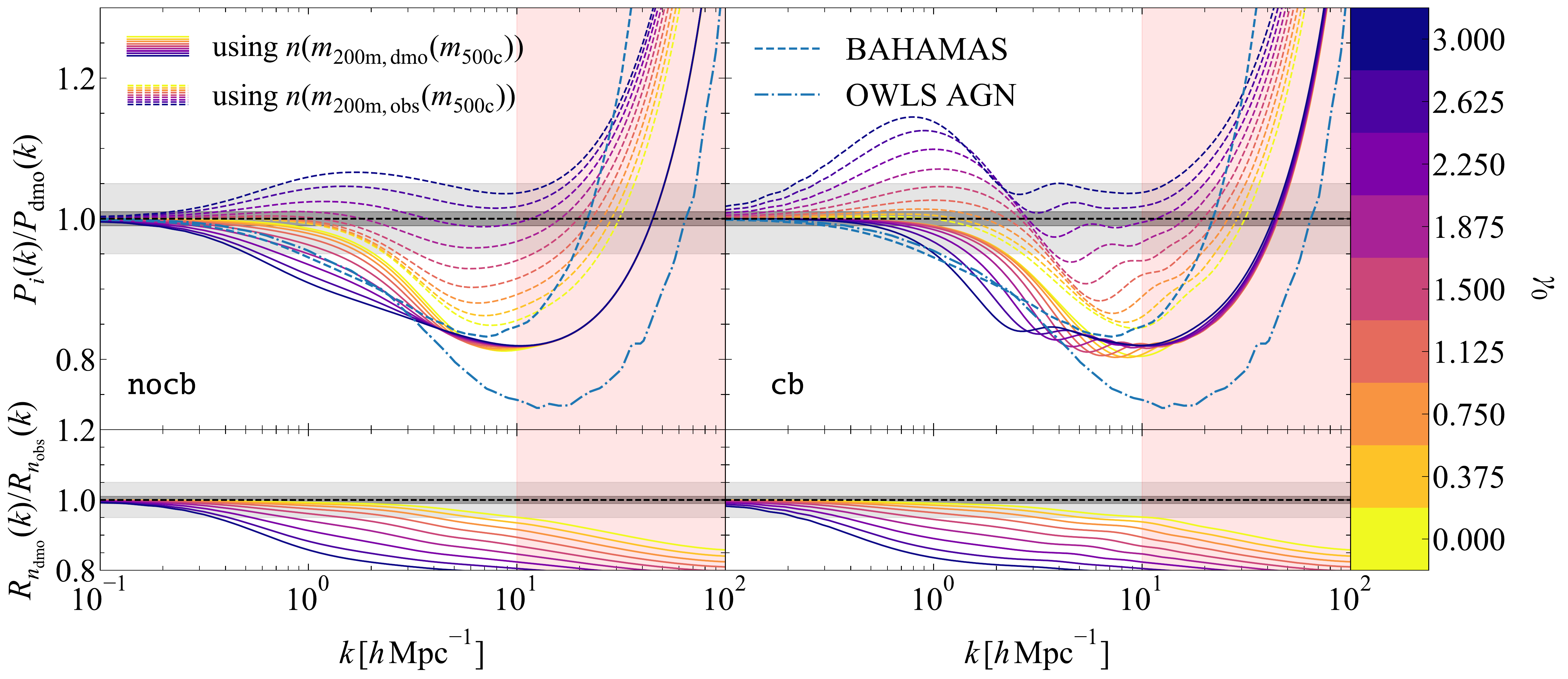}
  \caption{\emph{Top row} The influence of correcting the halo
    abundance for the halo mass decrease due to baryonic processes.
    The dark and light gray bands indicate the $\SI{1}{\percent}$ and
    $\SI{5}{\percent}$ intervals. The left-hand panel shows model
    \nocb, which effectively assumes that the baryons missing from
    haloes are redistributed far beyond $r\subscr{200m,obs}$ on linear
    scales. The right-hand panel shows model \cb, which adds the
    missing halo baryons in a uniform profile outside but near
    $r\subscr{200m,obs}$. The red, lightly-shaded region for
    $k > \SI{10}{\impch}$ indicates the scales where our model is is
    not a good indicator of the uncertainty because the stellar
    component is not varied. The lines are colour-coded by
    $\gamma_0 \equiv \gamma(m\subscr{500c} \to 0)$, the extrapolated
    power-law slope of the hot gas density profiles between
    $r\subscr{500c,obs}$ and $r\subscr{200m,obs}$, with lower values
    of $\gamma_0$ corresponding to flatter slopes. We show the power
    spectrum response with (our fiducial models, solid lines) and
    without (dashed lines) the correction for the halo masses applied.
    In the latter case, we find more power than in the DMO case at
    large scales because the abundance of low-mass haloes is
    overestimated due to not accounting for their mass loss compared
    to their DMO equivalents. \emph{Bottom row} The ratio of the power
    spectrum with and without correction for the halo masses. Models
    with flatter slopes, i.e. lower values of $\gamma_0$, reach the
    cosmic baryon fraction at $r\subscr{200m,obs}$ for lower halo
    masses, resulting in the same total mass-abundance relation as the
    DMO equivalent halo for a larger range of halo masses and thus
    more similar power spectra.}
  \label{fig:power_ratio_dndm}
\end{figure*}
Since halo abundances are generally obtained from N-body simulations,
it is crucial that we are able to correctly link observed haloes to
their dark matter only equivalents. However, astrophysical feedback
processes result in the ejection of gas and, consequently, a
modification of the halo profile and the halo mass $m\subscr{200m}$
\citep[e.g.][]{Sawala2013,Velliscig2014,Schaller2015}. Thus, not
accounting for the change in halo mass due to baryonic feedback would
result in the wrong relation between halo density profiles and halo
abundances in our model. Generally, feedback results in lower
extrapolated halo masses $m\subscr{200m,obs}$ for the observed haloes
than the DMO equivalent halo masses $m\subscr{200m,dmo}$. Thus, using
the observed mass instead of the DMO equivalent mass in the halo mass
function would result in an overprediction of the abundance of the
observed halo since $n(m)$ decreases with increasing halo mass.

We described how we link $m\subscr{200m,obs}$ to $m\subscr{200m,dmo}$
in \S~\ref{sec:hm_modifications}. We remind the reader that we assume
that baryons do not significantly alter the distribution of dark
matter. Thus, the dark matter component of the observed halo has the
same scale radius as its DMO equivalent and a mass that is a factor
$1-\Ob/\Om$ lower. The baryonic component of the observed halo is
determined by the observations and our different extrapolations for
$r>r\subscr{500c,obs}$. Then, from the total and rescaled DM density
profiles of the observed halo, we can determine the masses
$m\subscr{200m,obs}$ and $m\subscr{200m,dmo}$, respectively. These two
masses will differ because the baryons do not follow the dark matter.
The haloes have the abundance
$n(m\subscr{200m,dmo}(m\subscr{200m,obs}))$ for which we use the halo
mass function determined by \citet{Tinker2008}. In this section, we
test how this correction, i.e. using
$n(m\subscr{200m,dmo}(m\subscr{200m,obs}))$ instead of
$n(m\subscr{200m,obs})$, modifies our results.

We show the ratio of the observed halo mass to the DMO equivalent halo
mass at fixed radius $r\subscr{200c,dmo}$ in
Fig.~\ref{fig:mobs_r200c_dmo_m200c_dmo_ratio}. This ratio does not
depend on the model type, i.e. \cb\ or \nocb, since their density
profiles are the same for $r < r\subscr{200m,obs}$. We indicate the
range spanned by our models with $0 \leq \gamma_0 \leq 3$ by the red
shaded region. They converge at the high-mass end because not all
slopes $\gamma$ are allowed for high-mass haloes, as shown in
\S~\ref{sec:components}. At the low-mass end, our models converge
because the stellar component is fixed and hence does not depend on
$\gamma_0$, and the gas fractions approach 0. The thin, black, dotted
line indicates the ratio $1 - f\subscr{b}$ that our model converges to
when the halo baryon fraction reaches 0.

We also show the same relation found in the \owls\ AGN \citep[low-mass
haloes,][]{Schaye2010} and \cowls\ \citep[high-mass
haloes,][]{LeBrun2014a} simulations from \citet{Velliscig2014}. There
are systematic differences between the predictions from the
simulations and our model. These differences occur for two reasons.
First, our assumption that the baryons do not alter the distribution
of the dark matter with respect to the DMO equivalent halo, does not
hold in detail. \citet{Velliscig2014} show that at the fixed radius
$r\subscr{200c,dmo}$ there is a difference of up to $\SI{4}{\percent}$
between the dark matter mass of the observed halo and the dark matter
mass of the DMO equivalent halo, rescaled to account for the cosmic
baryon fraction. The dark matter in low-mass haloes expands due to
feedback expelling baryons outside $r\subscr{200c,dmo}$. In the
highest-mass haloes, feedback is less efficient and the dark matter
contracts in response to the cooling baryons. The thin, blue,
dash-dotted line shows the relation in \owls\ AGN when forcing the
dark matter mass of the halo to equal the rescaled DMO equivalent halo
mass. Hence, the contraction due to the presence of baryons of the
dark matter component for high-mass haloes explains the difference
between our model and the simulations. For the low-mass end, the
expansion of the dark matter component is not sufficient to explain
all of the difference. The remaining discrepancy results from the
higher baryons fractions in our model compared to the simulations.

We will neglect the response of the DM to the redistribution of
baryons throughout the rest of the paper. We have checked that scaling
the halo density profiles of the DMO equivalent haloes to match the
mass ratios from the \owls\ AGN simulation only affects our
predictions of the power suppression at the $\approx \SI{1}{\percent}$
level at our scales of interest (i.e. $k < \SI{10}{\impch}$). However,
even this small correction is an upper limit because we have assumed a
fixed ratio between the DM and rescaled DMO density profiles that
exceeds the correction for the cosmic baryon fraction. Hence, even at
large distances $r \gg r\subscr{200c,dmo}$ the mass ratio between the
halo and its DMO equivalent does not converge, whereas the mass
difference between hydrodynamical haloes and their DMO equivalents
eventually decreases to 0 \citep[see
e.g.][]{Velliscig2014,VanDaalen2014a}. We find such a small effect
because the low-mass haloes, whose mass ratio differs the most between
our model and the \owls\ AGN simulation, only have a small effect on
the total power at large scales, as we will show in
\S~\ref{sec:results_masses}.

We show how the correction of the halo abundance for the change in
halo mass due to baryonic processes affects the predicted power
spectrum response to baryons in Fig.~\ref{fig:power_ratio_dndm}. In
the top row, we show the power spectrum response for both the \nocb\
(left panel) and \cb\ models (right panel) with (our fiducial models,
solid lines) and without (dashed lines) the halo mass correction. When
not correcting the halo abundance for the change in halo mass (i.e.
when using $n(m\subscr{200m,obs})$), we actually find an increase in
power with respect to the DMO model at scales
$k \lesssim \SI{1}{\impch}$, i.e. $R_i(k) > 1$, for both the \nocb\
and \cb\ models, since the inferred abundances for observed haloes
with masses $m\subscr{200m,obs} \lesssim \SI{e14}{\mh}$ are too high.
At these scales, the Fourier profiles become constant in the 1h term,
i.e. $\hat{\rho}(k|m\subscr{h}) \to m\subscr{h}$ in Eq.~\ref{eq:p_1h},
and the power spectrum behaviour is thus dictated entirely by the halo
abundance. Hence, the power suppression that we find in our fiducial
models at these scales is the consequence of correcting the DMO
equivalent halo masses to account for the ejection of matter due to
feedback. We stress that our implementation of this effect is purely
empirical and does not rely on any assumptions about the physics
involved in baryonic feedback processes.

In the bottom row of Fig.~\ref{fig:power_ratio_dndm}, we show the
ratio between the power spectrum response to the inclusion of baryons
with and without the DMO equivalent halo mass correction. The
correction is most significant for the steepest extrapolated density
profile slopes, i.e. the highest values of $\gamma_0$, for which we
see the smallest ratios. For $\gamma_0 = 3$, even the most massive
haloes do not reach the cosmic baryon fraction inside
$r\subscr{200m,obs}$ (see Fig.~\ref{fig:fb_vs_m200m_obs}), and, hence,
even their abundances would be calculated wrongly if the observed
total mass was used, instead of the rescaled, observed DM mass, to
compute the halo abundance. In the case of the \cb\ models, there is
an extra increase in power for scales $k \lesssim \SI{1}{\impch}$ due
to the more extended baryon distribution.

It is striking how the halo mass correction modifies the suppression
of power in the way required to encompass the simulation predictions
at large scales. The correction to the DMO equivalent halo masses is
necessary for this match.

\subsection{Contribution of different halo
  masses}\label{sec:results_masses}
\begin{figure}
  \centering
  \includegraphics[width=\columnwidth]{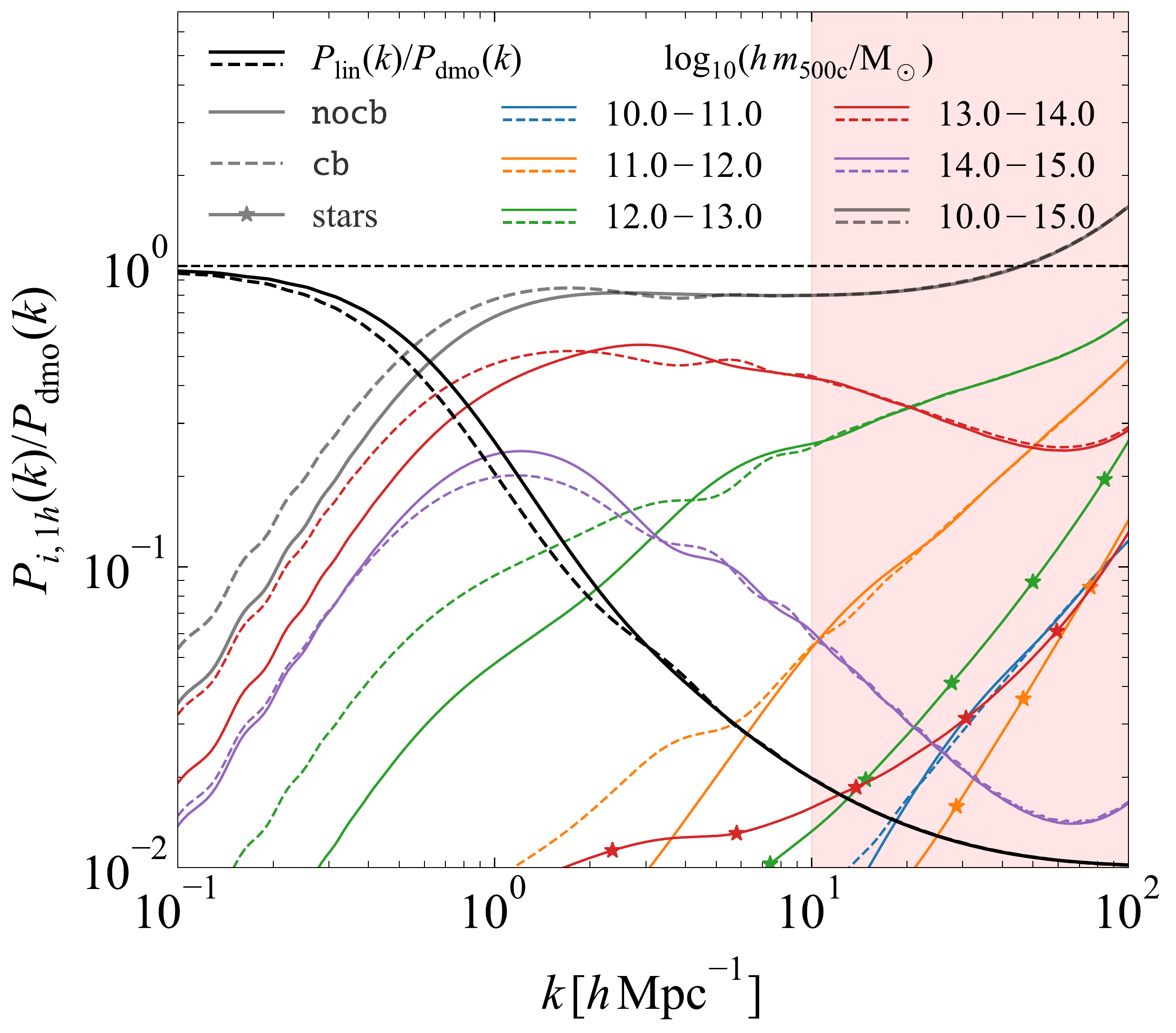}
  \caption{The contribution of the 1-halo term for different halo mass
    ranges to the total power spectrum at all scales for our fiducial
    models. The \nocb\ (solid lines) and \cb\ (dashed lines) models
    are shown and the mass ranges are indicated by the colours. We
    also show the contribution of the 2-halo term, i.e. the linear
    power spectrum (black lines). The stellar contribution (1h term
    and all cross-correlations between matter and stars, connected
    stars) is also included, but only for the \nocb\ case, since the
    \cb\ case traces the \nocb\ lines. The red, lightly-shaded region
    for $k > \SI{10}{\impch}$ indicates the scales where our model is
    is not a good indicator of the uncertainty because the stellar
    component is not varied. The 1h term dominates the total power for
    $k \gtrsim \SI{0.5}{\impch}$. For the scales of interest here
    ($k \lesssim \SI{5}{\impch}$), most of the power is contributed by
    groups and clusters with
    $\SI{e13}{\mh} \leq m\subscr{500c,obs} \leq \SI{e15}{\mh}$. For
    the \cb\ models, low-mass ($\approx \SI{e13}{\mh}$) haloes
    contribute more and clusters ($>\SI{e14}{\mh}$) less compared with
    the \nocb\ models. The total stellar contribution to the power
    response is $\lesssim \SI{1}{\percent}$ for all scales
    $k \lesssim \SI{5}{\impch}$ and only exceeds $\SI{2}{\percent}$
    for $k \gtrsim \SI{10}{\impch}$.}\label{fig:power_mass_contrib}
\end{figure}
To determine the observables that best constrain the matter power
spectrum at different scales, it is important to know which haloes
dominate the suppression of power at those scales. The dominant haloes
will be determined by the interplay between the total mass of the halo
and its abundance.

The halo model linearly adds the contributions from haloes of all
masses to the power at each scale. We show the contributions for five
decades in mass in Fig.~\ref{fig:power_mass_contrib} for our fiducial
model in the \nocb\ and \cb\ cases. We integrate the 1-halo term,
Eq.~\ref{eq:p_1h}, over 5 different decades in mass, spanning
$\SI{e10}{\mh} < m\subscr{500c,obs} < \SI{e15}{\mh}$, and then divide
each by the DMO power spectrum, showing the contribution of different
halo masses to the power spectrum. We also show the contribution of
the 2-halo term, i.e. the linear power spectrum. The mass dependence
of our model comes entirely from the 1h term, which dominates the
total power for $k \gtrsim \SI{0.5}{\impch}$.

We want to quantify the stellar contribution to the power spectrum to
gauge whether we are allowed to neglect the ISM component of the gas.
As explained in the beginning of \S~\ref{sec:hm_profiles}, we can
safely neglect the ISM if the stellar component contributes negligibly
to the total power at our scales of interest
($k \lesssim \SI{5}{\impch}$). To this end, we also include the 1h
term for the stellar component with all cross-correlations
$|\rho_\star(k|m\subscr{h})\rho_i(k|m\subscr{h})|$ in
Eq.~\eqref{eq:p_1h} with $i \in \{\mathrm{dm,gas}\}$. We only show
this contribution for the \nocb\ case, since the \cb\ results are
nearly identical. Fig.~\ref{fig:power_mass_contrib} clearly shows that
the stellar component contributes negligibly to the power for all
scales $k\lesssim \SI{5}{\impch}$ and, hence, we are justified in
neglecting the contributions of the ISM to the gas component. However,
for making predictions at the $\SI{1}{\percent}$ level on small scales
($k \gtrsim \SI{5}{\impch}$), the ISM and stellar components will
become important and will need to be modelled more accurately.

At all scales $k \lesssim \SI{10}{\impch}$, the total power is
dominated by groups
($\SI{e13}{\mh} \leq m\subscr{500c,obs} \leq \SI{e14}{\mh}$) and
clusters ($\SI{e14}{\mh} \leq m\subscr{500c,obs} \leq \SI{e15}{\mh}$)
of galaxies, with groups providing a similar or greater contribution
than clusters. Similar results have been found in DMO simulations by
\citet{VanDaalen2015}. Group-mass haloes have the largest range in
possible baryon fractions in our model, depending on the slope
$\gamma_0$ of the gas density profile for $r>r\subscr{500c,obs}$. We
conclude that groups are crucial contributors to the power at large
scales and thus measuring the baryon content of group-mass haloes will
provide the main observational constraint on predictions of the
baryonic suppression of the matter power spectrum.

\subsection{Influence of density profile fitting
  parameters}\label{sec:results_prms}
\begin{figure*}
  \centering
  \includegraphics[width=\textwidth]{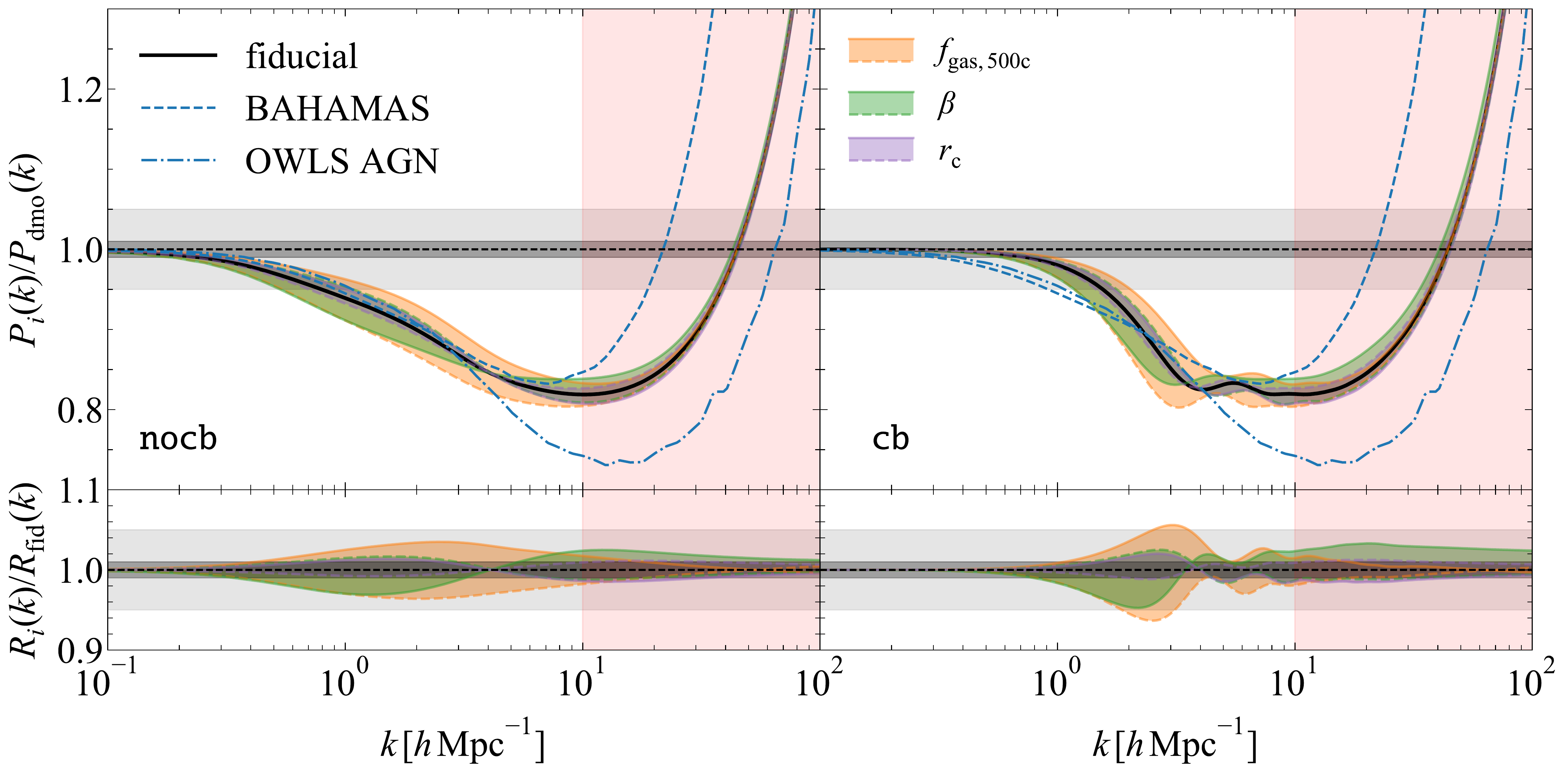}
  \caption{\emph{Top row} The variation in the power suppression due
    to baryonic effects when varying the best-fit hot gas density
    profile parameters independently within their $15\supscr{th}$
    (dashed lines) and $85\supscr{th}$ (solid lines) percentile ranges
    (shaded regions) and keeping all other parameters fixed. The dark
    and light gray bands indicate the $\SI{1}{\percent}$ and
    $\SI{5}{\percent}$ intervals. The left-hand panel shows model
    \nocb, which effectively assumes that the baryons missing from
    haloes are redistributed far beyond $r\subscr{200m,obs}$ on linear
    scales. The right-hand panel shows model \cb, which adds the
    missing halo baryons in a uniform profile outside but near
    $r\subscr{200m,obs}$. The red, lightly-shaded region for
    $k > \SI{10}{\impch}$ indicates the scales where our model is is
    not a good indicator of the uncertainty because the stellar
    component is not varied. The thick, solid, black line indicates
    our fiducial model with $\gamma_0 = 3\beta = 2.14$. \emph{Bottom
      row} The ratio between the region enclosed by the
    $15\supscr{th}-85\supscr{th}$ percentiles for each fit parameter
    and the fiducial model. The parameters $\beta$ and
    $f\subscr{gas,500c}$ determine the behaviour in the outer regions
    and are most important. The same trends are present for both the
    \nocb\ and \cb\ cases, but the \nocb\ case is sensitive to
    variations in the best-fit parameters out to larger scales. The
    uncertainty in any of the best-fit parameters allows at most a
    $\pm \SI{5}{\percent}$ variation in the power suppression for any
    scale.}\label{fig:power_var_prms}
\end{figure*}

So far, we have shown the impact of the baryon distribution and haloes
of different masses on the matter power spectrum for different
wavenumbers when assuming our model that best fits the observations.
However, since we assume the median values for the parameters
$r\subscr{c}$ and $\beta$, and the median relation
$f\subscr{gas,500c}-m\subscr{500c,obs}$ for the observed hot gas
density profiles and there is a significant scatter around these
medians, it is important to see how sensitive our predictions are to
variations in the parameter values. In this section, we investigate
the isolated effect of each observational parameter on the predicted
matter power spectrum.

We remind the reader of the beta profile in Eq.~\ref{eq:beta_gas} and
the best fits for its parameters determined from the observations in
Figs.~\ref{fig:obs_fgas}, \ref{fig:obs_rc_fit}, and
\ref{fig:obs_beta_fit}. In those figures, we indicated the median
relations, which are used in our model, and the $15\supscr{th}$ and
$85\supscr{th}$ percentiles of the observed values. We will test the
model response to variations in the hot gas observations by varying
each of the best-fit parameters between its $15\supscr{th}$ and
$85\supscr{th}$ percentiles while keeping all other parameters fixed.

We show the result of these parameter variations for our fiducial
model ($\gamma_0=2.14$) in the \nocb\ and \cb\ cases in
Fig.~\ref{fig:power_var_prms}. We indicate the $15\supscr{th}$
($85\supscr{th}$) percentile envelope with a dashed (solid), coloured
line and shade the region enclosed by these percentiles. For both the
\cb\ and \nocb\ cases, the parameters $\beta$ and $f\subscr{gas,500c}$
are the most important at large scales. Flatter outer slopes for the
hot gas density profile, i.e. smaller values of $\beta$, will result
in more baryons out to $r\subscr{200m,obs}$, yielding a smaller
suppression of power on large scales. Higher gas fractions within
$r\subscr{500c,obs}$ will result in haloes that are more massive and
contain more of the baryons, again yielding a smaller suppression of
power on large scales. The core radius $r\subscr{c}$ is the least
important parameter. Increasing the size of the core requires a lower
density in the core to reach the same gas fraction at
$r\subscr{500c,obs}$ and yields more baryons in the halo outskirts.
Hence, we see more power at large scales and less power on small
scales when increasing the value of $r\subscr{c}$ similarly to
decreasing the value $\beta$. However, the core is relatively close to
the cluster center and thus has no impact on the matter distribution
at large scales.

There is an important difference between the \nocb\ and \cb\ cases,
however. The fit parameters only start having an effect on the power
suppression at scales $k \gtrsim \SI{1}{\impch}$ in the \cb\ case,
whereas in the \nocb\ case they already start mattering around
$k \approx \SI{0.3}{\impch}$. If all baryons are accounted for in the
halo outskirts, as in the \cb\ case, the details of the baryon
distribution do not matter for the power at the largest scales, since
here the 1h term is fully determined by the mass inside $r\subscr{h}$,
which does not change for different values of $\beta$ and
$f\subscr{gas,500c}$. The \nocb\ model is a lot more sensitive to the
baryon distribution within the halo, since depending on the value of
$\beta$, or how many baryons can already be accounted for inside
$r\subscr{500c,obs}$, the haloes can have large variations in mass
$m\subscr{200m,obs}$.

In conclusion, the most important parameter to pin down is the gas
fraction of the halo, as we already concluded in
\S~\ref{sec:results_outer} and \S~\ref{sec:results_masses}. It has the
largest effect on all scales in both the \nocb\ and \cb\ cases and
varying its value within the observed scatter results in a
$\approx \pm \SI{5}{\percent}$ variation around the power spectrum
response predicted by our fiducial model. At the scales of interest to
future surveys, the effect of $\beta$ is of similar amplitude.
However, this parameter will be harder to constrain observationally
than the gas content of the halo, especially for group-mass haloes,
because X-ray observations cannot provide an unbiased sample and SZ
observations cannot observe the density profile directly.

\subsection{Influence of hydrostatic bias}\label{sec:results_bias}
\begin{figure*}
  \centering
  \includegraphics[width=\textwidth]{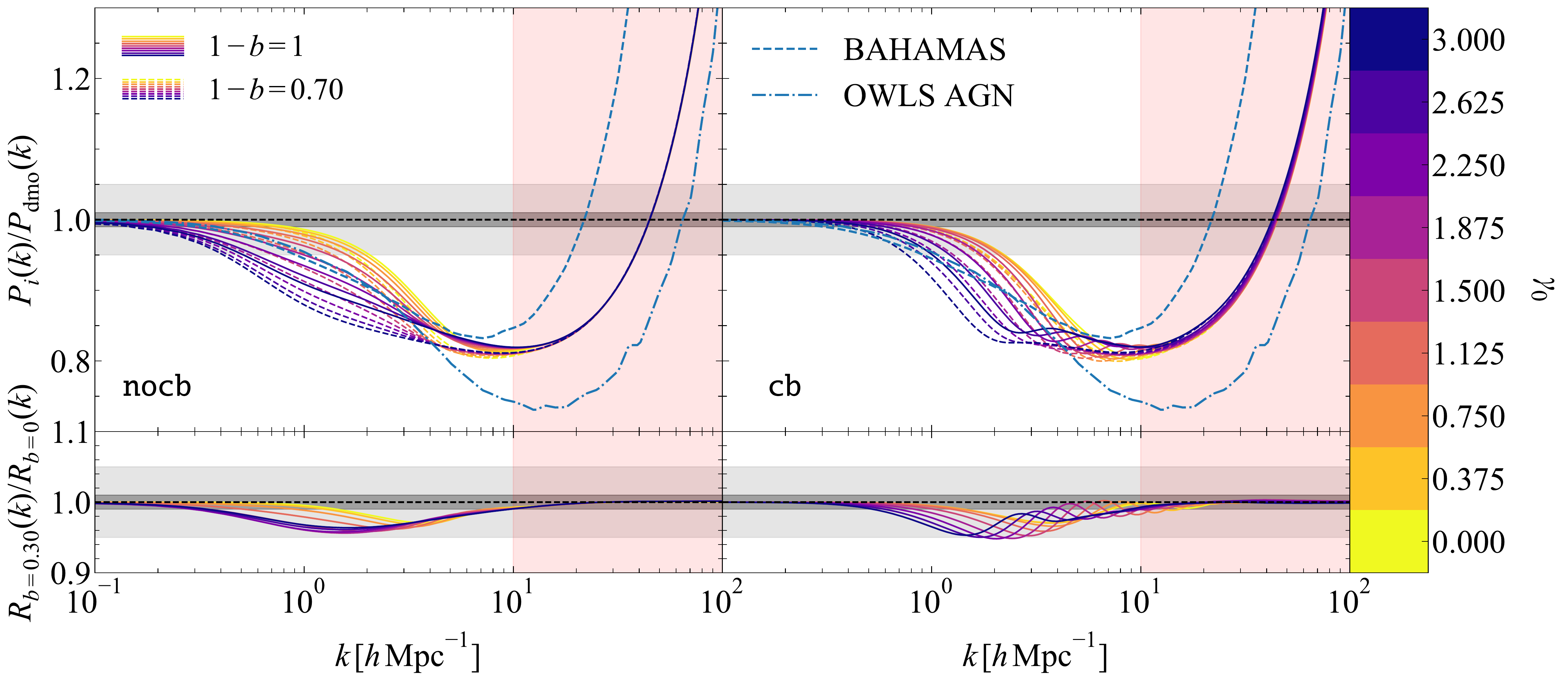}
  \caption{\emph{Top row} The response of the matter power spectrum to
    the presence of baryons with (dashed lines) and without (solid
    lines) accounting for a hydrostatic bias of $1-b=0.7$ in the halo
    mass. The dark and light gray bands indicate the
    $\SI{1}{\percent}$ and $\SI{5}{\percent}$ intervals. The left-hand
    panel shows model \nocb, which effectively assumes that the
    baryons missing from haloes are redistributed far beyond
    $r\subscr{200m,obs}$ on linear scales. The right-hand panel shows
    model \cb, which adds the missing halo baryons in a uniform
    profile outside but near $r\subscr{200m,obs}$. The red,
    lightly-shaded region for $k > \SI{10}{\impch}$ indicates the
    scales where our model is is not a good indicator of the
    uncertainty because the stellar component is not varied.
    \emph{Bottom row} The ratio between the corrected and the
    uncorrected models. Haloes with lower values of $\gamma_0$ are
    less strongly affected by the bias since they can add more baryons
    outside $r\subscr{500c,obs}$. Without correcting for the bias, we
    underestimate the suppression of power by up to
    $\approx \SI{4}{\percent}$ at
    $k = \SI{1}{\impch}$.}\label{fig:power_ratio_debiased}
\end{figure*}
All of our results so far assumed gas fractions based on halo masses
derived from X-ray observations under the assumption of hydrostatic
equilibrium (HE) and pure thermal pressure. Under the HE assumption
non-thermal pressure and large-scale gas motions are neglected in the
Euler equation \citep[see e.g. the discussion in \S 2.3
of][]{Pratt2019}. However, in massive systems in the process of
assembly, there is no a priori reason to assume that simplifying
assumption to hold. We expect the most massive clusters to depart from
HE, since we know from the hierarchical structure formation paradigm,
that they have only recently formed. Moreover, the pressure can have a
non-negligible contribution from non-thermal sources such as
turbulence \citep{Eckert2018}.

Investigating the relation between hydrostatically derived halo masses
and the true halo mass requires hydrodynamical simulations
\citep[e.g.][]{Nagai2007, Rasia2012, Biffi2016, Brun2017,
  McCarthy2017, Henson2017} or weak gravitational lensing observations
\citep{Mahdavi2013, VonderLinden2014, Hoekstra2015, Medezinski2018}.
In both cases, the pressure profile of the halo is derived from
observations of the hot gas. Under the assumption of spherical
symmetry and HE, this pressure profile is then straightforwardly
related to the total mass profile of the halo. Subsequently, this
hydrostatic halo mass can be compared to an unbiased estimate of the
halo mass, i.e. the true mass in hydrodynamical simulations, or the
mass derived from weak lensing observations.

The picture arising from both simulations and observations is that
hydrostatic masses, $m\subscr{HE}$, are generally biased low with
respect to the weak lensing or true halo mass, $m\subscr{WL}$, with
$m\subscr{HE}/m\subscr{WL} = 1-b \simeq \numrange{0.6}{0.9}$ (e.g.
\citealp{Mahdavi2013}; \citealp{VonderLinden2014};
\citealp{Hoekstra2015}; \citealp{Brun2017}; \citealp{Henson2017};
\citealp{Medezinski2018}). The detailed behaviour of this bias depends
on the deprojected temperature and density profiles, with more
spherical systems being less biased.

Correcting for the observationally determined bias would result in
higher halo masses and, consequently, a shift in the gas fractions
away from the assumed best-fit $f\subscr{gas,500c}-m\subscr{500c,obs}$
relation. We argued previously that this is the most relevant
observable to determine the suppression of power at scales
$k \lesssim \SI{1}{\impch}$. Thus, it is important to investigate how
the HE assumption affects our predictions. Previously,
\citet{Schneider2018} have shown for three different levels of
hydrostatic bias ($1 - b \in \{0.71, 0.83, 1\}$) that the predicted
power suppression at large scales $k < \SI{1}{\impch}$ can vary by up
to $\SI{5}{\percent}$.

Staying in tune with \S~\ref{sec:results_prms}, we adopt a single
value for the bias to investigate its influence on our predictions. We
will take $1-b=0.7$ which is consistent with both
\citet{VonderLinden2014} and \citet{Hoekstra2015}. Moreover, although
the bias tends to be higher for higher-mass systems because of the
presence of cooler gas in their outskirts \citep{Henson2017}, we
conservatively adopt this value for all halo masses. Correcting for
the bias will influence our model in two ways. First, the inferred gas
masses will increase slightly, since the true $r\subscr{500c,obs}$
will be larger than the value assumed from the hydrostatic estimate.
We thus recompute the gas masses from our best-fit beta models to the
observations. Second, the halo mass will increase by the bias factor
which will result in new estimates for the gas fractions, which we
show as the thin, solid, black line in Fig.~\ref{fig:obs_fgas}. We
then fit the median $f\subscr{gas,500c}-m\subscr{500c,obs}$ relation
again, assuming Eq.~\ref{eq:fgas_sigmoid}, resulting in the thin, red,
dashed line.

We show the resulting effect on the baryonic suppression of the power
spectra in Fig.~\ref{fig:power_ratio_debiased}. The results are
similar to varying $f\subscr{gas,500c}$ in
Fig.~\ref{fig:power_var_prms}, since the bias-corrected relation is
similar to the 15\supscr{th} percentile
$f\subscr{gas,500c}-m\subscr{500c,obs}$ relation, but with a more
dramatic suppression of the baryonic mass for clusters and hence more
suppression of the power at large scales. In the bottom panels of
Fig.~\ref{fig:power_ratio_debiased}, we find a maximum extra
suppression of $\approx \SI{4}{\percent}$ due to the hydrostatic bias
at $k = \SI{1}{\impch}$ in both the \nocb\ and \cb\ cases, which is
consistent with the findings of \citet{Schneider2018}. The magnitude
of the suppression is lower for lower values of $\gamma_0$ since these
models compensate for the lower baryon fraction within
$r\subscr{500c,obs}$ by adding baryons between $r\subscr{500c,obs}$
and $r\subscr{h}$.

Accounting for the bias breaks the overall agreement with the
simulations on large scales for the models with high values of
$\gamma_0$. However, in \bah\ and \owls\ AGN, a hydrostatic bias of
$1-b = 0.84$ and $1-b=0.8$ is found, respectively, for groups and
clusters \citep{McCarthy2017,LeBrun2014a}. When we assume $1-b=0.8$,
we find a maximum extra suppression of $\approx \SI{2}{\percent}$ at
$k = \SI{1}{\impch}$ instead of $\approx \SI{4}{\percent}$. At other
scales the effect of the hydrostatic bias is similarly reduced.

In conclusion, it is crucial to obtain robust constraints on the
hydrostatic bias of groups and clusters of galaxies. Current
measurements of this bias suggest that hydrostatic halo masses
underestimate the true masses and that this bias results in a downward
shift of the cluster gas fractions that is more severe than the
observational scatter in the relation. Because the shift affects
cluster-mass haloes, it results in an additional power suppression of
up to $\approx \SI{4}{\percent}$ at $k = \SI{1}{\impch}$, depending on
how our model distributes the outer baryons. There are ways of
measuring halo masses that do not rely on making the hydrostatic
assumption, such as weak lensing observations, but these also carry
their own intrinsic biases \citep{Henson2017}. Making mock
observations in simulations allows us to characterize these separate
biases \citep[e.g.][]{Henson2017,Brun2017}, but the simulations still
do not make a full like-for-like comparison with the observations.
Finally, joint constraints on X-ray, SZ, and weak lensing halo mass
scaling relations, including possible biases, as was done in
\citet{Bocquet2019}, could provide more robust halo mass estimates.

\section{Discussion}\label{sec:discussion}
We have presented an observationally constrained halo model to
estimate the power suppression due to baryons without any reliance on
subgrid recipes for the unresolved physics of baryons in
hydrodynamical simulations. We reiterate that our main goal is not to
provide the most accurate predictions of the matter power spectrum,
but to investigate the possibility of using observations to constrain
it. The fact that the clustering of matched haloes does not change
between DMO and hydrodynamical simulations \citep{VanDaalen2014a}
implies that changes in the density profiles due to the baryons
determine the change of the matter power spectrum. Hence, even though
the halo model does not accurately predict the matter power spectrum,
it can accurately predict the relative effect of baryonic processes on
the power spectrum. The overall agreement between our model and
hydrodynamical simulations that reproduce the observed distribution of
baryons in groups and clusters, confirms that our model captures the
first-order impact of baryons simply by reproducing the observed
baryon content for groups and clusters.

In conclusion, the main strength of the model is that it allows us to
quantify the impact of different halo masses, different halo baryon
density distributions and observational biases and uncertainties on
the baryonic suppression of the matter power spectrum without any
necessity for uncertain subgrid recipes for feedback processes. This
in turn allows us to provide a less model-dependent estimate of the
range of possible baryonic suppression and to predict which
observations would provide the strongest constraints on the matter
power spectrum.

There are other models in the literature that aim to model the effect
of baryon physics on the matter power spectrum. \texttt{HMcode} by
\citet{Mead2015} is widely used to include baryon effects in weak
lensing analyses. Although \texttt{HMcode} is also based on the halo
model, its aim is different from ours. \citet{Mead2015} modify the
dark matter halo profiles and subsequently fit the parameters of their
halo model to hydrodynamical simulations to provide predictions for
the baryonic response of the power spectrum that are accurate at the
$\sim \SI{5}{\percent}$ level for $k \lesssim \SI{5}{\impch}$ with 2
free parameters related to the baryonic feedback (for a similar
approach, see \citealp{Semboloni2013}). These feedback parameters can
then be jointly constrained with the cosmology using cosmic shear
data. However, even though the modifications to the dark matter
profile are phenomenologically inspired, there is no guarantee that
the final best-fit parameters correspond to the actual physical state
of the haloes. We obtain similar accuracy in the predicted power
response when viewing $\gamma_0$ as a fitting parameter and comparing
to hydrodynamical simulations. However, in our case, fitting
$\gamma_0$ preserves the agreement with observations. Indeed, the most
important difference between our approach and that of \citet{Mead2015}
is that we fit to observations instead of simulations.

The investigation of \citet{Schneider2018} most closely matches our
goal. \citet{Schneider2015} and \citet{Schneider2018} developed a
\emph{baryon correction model} to investigate the influence of baryon
physics on the matter power spectrum. Their model shifts particles in
DMO simulations according to the physical expectations from baryonic
feedback processes. Since the model only relies on DMO simulations, it
is not as computationally expensive as models that require
hydrodynamical simulations to calibrate their predictions. Our simple
analytic halo model is cheaper still to run, but it only results in a
statistical description of the matter distribution, whereas the
\emph{baryon correction model} predicts the total matter density field
for the particular realization that was simulated. Because our model
combines the universal DMO halo mass function with observed density
profiles, it can easily be applied to a wide variety of cosmologies
without having to run an expensive grid of DMO simulations.

In the baryon correction model, the link to observations can also be
made, making it similar to our approach. \citet{Schneider2018} fit a
mass-dependent slope of the gas profile, $\beta$ (note that their
slope is not defined the same way as our slope $\beta$), and the
maximum gas ejection radius, $\theta\subscr{ej}$, to the observed hot
gas profiles of the XXL sample of \citet{Eckert2016} and a compendium
of X-ray gas fraction measurements. They also include a stellar
component that is fit to abundance matching results, similar to our
\texttt{iHOD} implementation. They show that their model can reproduce
the observed relations as well as hydrodynamical simulations when fit
to their gas fractions. \citet{Schneider2018} use the observations to
set a maximum range on their model parameters to then predict both the
matter power spectrum and the shear correlation function. Our work, on
the other hand, focusses on the impact of isolated properties of the
baryon distribution on the power spectrum. Similarly to
\citet{Schneider2018}, we find that the power suppression on large
scales is very sensitive to the baryon distribution in the outskirts
of the halo. However, our model allows us to clearly show that the
halo baryon fractions are the crucial ingredient in setting the total
power suppression at large scales, $k \lesssim \SI{1}{\impch}$. Also
similarly to \citet{Schneider2018}, we find that the hydrostatic mass
bias significantly affects the total power suppression at large
scales.

So far, we have not included redshift evolution. \citet{Schneider2018}
have found that the most important evolution of clusters and groups in
cosmological simulations stems from the change in their abundance due
to the evolution of the halo mass function in time, and not due to the
change of the density profiles with time. This evolution can be
readily implemented into our halo model.

\section{Summary and conclusions}\label{sec:summary_conclusions}

Future weak lensing surveys will be limited in their accuracy by how
well we can predict the matter power spectrum on small scales
\citep[e.g.][]{Semboloni2011,Copeland2018,Huang2019}. These scales
contain a wealth of information about the underlying cosmology of our
Universe, but the interpretation of the signal is complicated by
baryon effects. Our current theoretical understanding of the impact of
baryons on the matter power spectrum stems from hydrodynamical
simulations that employ uncertain subgrid recipes to model
astrophysical feedback processes. This uncertainty can be bypassed by
adopting an observational approach to link the observed distribution
of matter to the matter power spectrum.

We have provided a detailed study of the constraints that current
observations of groups and clusters of galaxies impose on the possible
influence of the baryon distribution on the matter power spectrum. We
introduced a modified halo model that includes dark matter, hot gas,
and stellar components. We fit the hot gas to X-ray observations of
clusters of galaxies and we assumed different distributions for the
missing baryons outside $r\subscr{500c,obs}$, the maximum radius
probed by X-ray observations of the hot gas distribution.
Subsequently, we quantified (i) how the outer, unobserved baryon
distribution modifies the matter power spectrum
(Fig.~\ref{fig:power_ratio}). We also investigated (ii) how the change
in halo mass due to baryonic effects can be incorporated into the halo
model (Fig.~\ref{fig:power_ratio_dndm}). We showed (iii) the
contributions to the matter power spectrum of haloes of different
masses at different spatial scales
(Fig.~\ref{fig:power_mass_contrib}), (iv) the influence of varying the
individual best-fit parameters to the observed density profiles within
their allowed range (Fig.~\ref{fig:power_var_prms}), and (v) the
influence of a hydrostatic mass bias on the matter power spectrum
(Fig.~\ref{fig:power_ratio_debiased}).

Our model has one free parameter, $\gamma_0$, related to the slope of
the hot gas density profile for
$r\subscr{500c,obs} \leq r \leq r\subscr{200m,obs}$, where
observational constraints are very poor. We considered two extreme
cases for the baryons. First, the \nocb\ models assume that haloes of
size $r\subscr{200m,obs}$ do not necessarily reach the cosmic baryon
fraction at this radius and that any missing baryons are located at
such large distances that they only contribute to the 2-halo term.
Second, in the \cb\ models the missing baryons inside
$r\subscr{200m,obs}$ are distributed with an assumed uniform density
profile outside this radius until the cosmic baryon fraction is
reached. These cases provide, respectively, the maximum and minimum
power suppression of large-scale power due to baryonic effects.

All of our observationally constrained models predict a significant
amount of suppression on the scales of interest to future surveys
($\num{0.2} \lesssim k / (\si{\impch}) \lesssim \num{5}$). We find a
total suppression of $\SI{1}{\percent}$ ($\SI{5}{\percent}$) on scales
$\SIrange{0.2}{0.9}{\impch}$ ($\SIrange{0.5}{2}{\impch}$) in the
\nocb\ case and on scales $\SIrange{0.5}{1}{\impch}$
($\SIrange{1}{2}{\impch}$) in the \cb\ case for values
$\gamma_0 = \numrange{3}{0}$ (Fig.~\ref{fig:power_ratio}), where
$\gamma_0$ is the low-mass limit of the power-law slope $\gamma$
between $r\subscr{500c,obs}$ and $r\subscr{200m,obs}$, i.e.
$\gamma_0 = \gamma(m\subscr{500c,obs} \to 0)$. This large possible
range of scales corresponding to a fixed suppression factor for each
case illustrates the importance of the baryon distribution outside
$r\subscr{500c,obs}$ (which is parameterised by $\gamma_0$) in setting
the total power suppression.

We found that massive groups of galaxies
($\SI{e13}{\mh} < m\subscr{500c,obs} < \SI{e14}{\mh}$) provide a
larger contribution than clusters to the total power at all scales
(Fig.~\ref{fig:power_mass_contrib}). This is unfortunate, since we
have shown that the baryonic content of group- and cluster-sized
haloes, which is set by the observed gas fractions
$f\subscr{gas,500c}$, determines the large-scale
($k \lesssim \SI{1}{\impch}$) power suppression
(Figs.~\ref{fig:power_vs_fbar} and~\ref{fig:power_var_prms}). However,
observations of the hot gas content of groups are scarcer than those
of clusters and are also subject to a considerable Malmquist bias.
Current X-ray telescopes cannot solve this problem, but a combined
approach with Sunyaev-Zel'dovich or gravitational lensing observations
could provide a larger sample of lower mass objects.

We found that our observationally constrained models only encompass
the predictions of hydrodynamical simulations that reproduce the hot
gas content of groups and clusters of galaxies
(Fig.~\ref{fig:power_ratio_sims}). Thus, we stress the importance of
using simulations that reproduce the relevant observations when using
such models to predict the baryonic effects on the matter
distribution.

We found that accurately measuring the halo masses is of vital
importance when trying to place observational constraints on the
matter power spectrum. An unrecognized hydrostatic halo mass bias of
$1-b=0.7$ would result in an underestimate of the total power
suppression by as much as $\SI{4}{\percent}$ at $k = \SI{1}{\impch}$
(Fig.~\ref{fig:power_ratio_debiased}). In addition, it is critical to
correct the observed halo masses for the redistribution of baryons
when estimating their abundance using halo mass functions based on DM
only simulations (Fig.~\ref{fig:power_ratio_dndm}).

All in all, it is encouraging that we are able to quantify the
baryonic suppression of the matter power spectrum with a simple,
flexible but physical approach such as our modified halo model. Our
investigation allows us to predict the observations that will be most
constraining for the impact of baryonic effects on the matter power
spectrum.

\section*{Acknowledgements}

It is a pleasure to thank Ian McCarthy and Marcel van Daalen for
useful discussions and comments. We would also like to thank Ian
McCarthy and Dominique Eckert for providing the observational data and
the REXCESS team for making their density profiles publically
available. Finally, we would like to thank the referee for a
constructive report that helped clarify the paper. The authors
acknowledge support from: the Netherlands Organisation for Scientific
Research (NWO) under grant numbers 639.043.512 (SD, HH) and
639.043.409 (SD, JS).



\bibliographystyle{mnras}
\bibliography{halo_model.bbl}



\appendix

\section{Influence of the halo mass
  range}\label{app:results_mass_range}
In this section, we investigate how our choice of mass grid influences
our predictions. We have chosen an equidistant log-grid of halo masses
\(\SI{e11}{\mh} \leq m\subscr{500c,obs} \leq \SI{e15}{\mh}\), sampled
with 101 bins. Doubling or halving the number of bins only affects our
predictions at the $< \SI{0.1}{\percent}$ level for all $k$.
Similarly, increasing the maximum halo mass to
$m\subscr{500c,max} = \SI{e16}{\mh}$ only results in changes at the
$< \SI{0.1}{\percent}$ level for all $k$. The only significant change
occurs when decreasing the minimum halo mass to
$m\subscr{500c,min} = \SI{e6}{\mh}$, but this only affects scales
smaller than of interest here. In this case, our baryonic models
predict less power compared to the higher minimum mass case, since the
low-mass haloes have no stars and gas. Hence, they will always contain
less matter than their DMO equivalents and the DMO power will be
boosted relative to the baryonic one. However, our predictions only
change at the $\SI{1}{\percent}$ level for
$k \gtrsim \SI{60}{\impch}$, thus our fiducial mass range is converged
for our scales of interest, $k < \SI{10}{\impch}$.

\section{Influence of concentration-mass relation}\label{app:results_concentration}
In this section, we investigate how changes in the concentration at
fixed halo mass influences our predictions. While the
concentration-mass relation does not show a strong mass dependence,
the scatter about the median relation is significant \citep{Jing2000,
  Bullock2001, Duffy2008, Dutton2014}. To investigate the potential
influence of this scatter, we tested how our predictions for the power
response due to baryons change when assuming the $c(m)$ relation
shifted up and down by its log-normal scatter
$\sigma_{\log_{10} c} = 0.15$ \citep{Duffy2008, Dutton2014}.
Increasing the concentration results in more (less) power at small
(large) scales and thus a lower (higher) power suppression. Adopting
this extreme shift in the concentration-mass relation results in a
maximum variation of $\pm \SI{3}{\percent}$ in suppression at scales
$k \lesssim \SI{20}{\impch}$. This variation is smaller than any of
the hot gas density profile best-fit parameter variations in
\S~\ref{sec:results_prms}.

\bsp	
\label{lastpage}
\end{document}